\documentclass[11pt]{article}
\usepackage{jheppub} 
\usepackage{physics}
\usepackage{amsmath}
\usepackage{amsthm}
\usepackage{slashed}
\usepackage[dvipsnames]{xcolor}
\usepackage{mathtools}
\usepackage{bbold}
\usepackage{marvosym}
\usepackage{ytableau}
\usepackage{mathdots}
\usepackage{dynkin-diagrams}
\usepackage{enumitem}
\usepackage{slashed}
\usepackage{rotating}
\usepackage[compat=1.1.0]{tikz-feynman}
\usepackage{graphicx}
\usepackage{caption}
\usepackage{subcaption}
\usepackage{arydshln}

\renewcommand{\vev}[1]{\langle #1 \rangle}

\title{Electroweak-Charged Dark Matter and \boldmath$SO(10)$ Unification with Parity}
\author[a,b]{Matthew J. Baldwin}

\author[a,b,c]{and Keisuke Harigaya}

\affiliation[a]{Department of Physics, University of Chicago, Chicago, IL 60637, USA}
\affiliation[b]{Enrico Fermi Institute and Kavli Institute for Cosmological Physics, University of Chicago, Chicago, IL 60637, USA}
\affiliation[c]{Kavli Institute for the Physics and Mathematics of the Universe (WPI),
The University of Tokyo Institutes for Advanced Study,
The University of Tokyo, Kashiwa, Chiba 277-8583, Japan}

\emailAdd{mjbaldwin@uchicago.edu}
\emailAdd{kharigaya@uchicago.edu}

\abstract{
We consider electroweak-charged dark matter in an $SO(10)$ unified theory that solves the strong $CP$ problem via Parity. Electroweak-charged dark matter has a colored $SO(10)$ partner, whose mass should be much above the dark matter mass to avoid cosmological problems arising from the decay of the colored partner. The mass hierarchy can be naturally achieved by an $SO(10)\times CP$ symmetry breaking Higgs that has a missing vacuum expectation value. The mass hierarchy, via quantum corrections to the gauge coupling constants, lowers the unification scale and enhances the proton decay rate. Hyper-Kamiokande will probe the parameter space with precise gauge coupling unification. We derive the range of the top quark mass and the strong coupling constant preferred by radiative Parity breaking by the Higgs Parity mechanism.
}

\begin{document}
\maketitle
\flushbottom
\newpage

\section{Introduction}
The symmetry structure of the Standard Model (SM) of particle physics remains mysterious. The weak interaction violates $CP$ symmetry through the Yukawa couplings of the quarks, which is expected to induce $CP$ violation in the strong interaction~\cite{Bell:1969ts,Adler:1969gk,tHooft:1976rip,tHooft:1976snw}. However, the magnitude of $CP$ violation in the strong interaction is smaller than the naive expectation by more than ten orders of magnitude~\cite{nEDM:2020crw}.

The absence of strong $CP$ violation may be explained by a spontaneously broken discrete space-time symmetry. We focus on Parity symmetric models~\cite{Beg:1978mt,Mohapatra:1978fy,Babu:1988mw,Babu:1989rb,Kuchimanchi:1995rp,Mohapatra:1995xd,Hall:2018let}, where the gauge group of the SM is extended to $SU(3)_c\times SU(2)_L\times SU(2)_R\times U(1)_X$ and Parity exchanges $SU(2)_L$ with $SU(2)_R$. A crucial advantage of Parity symmetry over $CP$ symmetry in solving the strong $CP$ problem is that the Yukawa couplings are only required to be Hermitian, rather than real, and weak $CP$ violation is readily obtained. See~\cite{Barr:1991qx,Bonnefoy:2023afx} for Parity symmetric models with different gauge groups and \cite{Nelson:1983zb,Barr:1984qx,Bento:1991ez,Hiller:2001qg,Vecchi:2014hpa,Dine:2015jga,Girmohanta:2022giy,Kuchimanchi:2023imj} for CP symmetric models. 

There are also several phenomenological advantages of Parity. Parity requires right-handed neutrinos, whose coupling to SM neutrinos can give Majorana neutrino masses through the seesaw mechanism~\cite{Yanagida:1979as,GellMann:1980vs,Minkowski:1977sc,Mohapatra:1979ia,Hall:2023vjb} or Dirac neutrino masses radiatively~\cite{Babu:1988yq,Babu:2022ikf}. Out-of-equilibrium decay of the right-handed neutrinos can explain the observed baryon asymmetry~\cite{Fukugita:1986hr,Buchmuller:2004nz,Giudice:2003jh}. The extended gauge group can be embedded into the $SO(10)$ grand unified group, and precise gauge coupling unification fixes the possible range of the Parity symmetry breaking scale~\cite{Gipson:1984aj,Chang:1984qr,Deshpande:1992au,Bertolini:2009qj,Hall:2019qwx,Hamada:2020isl}.

One of the right-handed neutrinos can in principle be dark matter~\cite{Bezrukov:2009th,Dror:2020jzy}. However, enough stability of dark matter requires that the dark matter right-handed neutrino have only very small Yukawa couplings. In $SO(10)$ theories, the Yukawa couplings of right-handed neutrinos are related with up-type Yukawa couplings and it is challenging to make the right-handed neutrino stable enough.

In this paper, we instead introduce electroweak-charged dark matter in an $SO(10)$ unified theory. Electroweak-charged dark matter is phenomenologically interesting. Assuming that the reheating temperature of the universe is above the dark matter mass, the abundance of dark matter is determined by the freeze-out mechanism~\cite{Lee:1977ua} and the mass of dark matter is predicted to be around the TeV scale. We may detect dark matter directly by nucleon recoil experiments and/or indirectly by the observations of cosmic rays.

The existence of electroweak-charged dark matter can affect gauge coupling unification. In grand unified theories, electroweak-charged dark matter has colored partners that decay into dark matter and SM particles via the exchange of heavy gauge bosons with masses around the unification scale. The colored partners should be much heavier than dark matter. If not, the colored partners are long-lived and may overproduce dark matter or disturb Big-Bang Nucleosynthesis (BBN). The required mass splitting can be naturally obtained by the missing vacuum expectation value (VEV) structure of an $SO(10)\times CP$ breaking Higgs. The mass splitting changes the running of the gauge coupling constants and affects the prediction on the Parity breaking scale and the unification scale. We find that the Parity breaking scale becomes higher and the unification scale becomes lower compared to the case without dark matter. This makes the observation of proton decay in the near future more likely. We find that Hyper-Kamiokande can probe parameter space with precise gauge coupling unification with $\Delta < 7$, where $\Delta \sim {\rm max}_{i,j=1,2,3}|2\pi/\alpha_i-2\pi/\alpha_j|$.

Our results also have implications to the measurements of SM parameters. In the minimal Higgs model~\cite{Babu:1988mw,Babu:1989rb}, the SM Higgs quartic coupling is predicted to nearly vanish at the Parity symmetry breaking scale~\cite{Hall:2018let}, so precise gauge coupling unification predicts the values of the SM parameters, particularly the top quark mass and the strong coupling constant. We derive this prediction in the $SO(10)$ model with electroweak-charged dark matter. Such correlations between beyond-SM and SM parameters have been studied for models of baryogenesis~\cite{Dunsky:2020dhn,Carrasco-Martinez:2023nit} and dark matter~\cite{Dunsky:2019api,Dunsky:2019upk,Dror:2020jzy}.

The connection between electroweak-charged dark matter and gauge coupling unification has also been discussed in the literature. Refs.~\cite{Ibe:2009gt,Aizawa:2014iea,Harigaya:2016vda} consider $SU(5)$ unification with split dark matter multiplets. Refs.~\cite{Mambrini:2015vna,Nagata:2015dma} consider $SO(10)$ unification with split dark matter multiplets and intermediate gauge symmetry breaking. 

This paper is organized as follows. In Sec.~\ref{sec:unification}, we review $SO(10)$ unification with Parity symmetry and how the strong $CP$ problem is solved. Sec.~\ref{sec:WIMP_DM} discusses the cosmological constraints on electroweak-charged dark matter candidates in the $SO(10)$ theory and how the required mass splitting can be achieved. In Sec.~\ref{sec:gauge_unifcation}, we compute the running of the gauge couplings and matching conditions to discuss the quality of unification, compute the proton decay rate, and provide constraints on the Parity symmetry breaking scale. The predictions on SM parameters are given in Sec.~\ref{sec:SM}.

\section{Parity and \boldmath$SO(10)$ Unification}
\label{sec:unification}

In this section, we review $SO(10)$ unification with a spontaneously broken Parity symmetry developed in~\cite{Hall:2019qwx}. We first discuss $SO(10)$ breaking down to $SU(3)_c\times SU(2)_L\times SU(2)_R\times U(1)_X$ ($\equiv G_{LR}$). We then discuss $G_{LR}$ breaking down to $SU(3)_c\times SU(2)_L\times U(1)_Y$ ($\equiv G_{SM}$) and show how the Parity symmetry breaking scale is correlated with the SM Higgs quartic coupling. Finally, we show how the SM Yukawa couplings are obtained and the strong $CP$ problem is solved.

\subsection{$SO(10)$ breaking}\label{sec:SO(10)_breaking}

$SO(10)\times CP$ symmetry is broken by a non-zero VEV of a $CP$-odd Higgs in the ${\bf 45}$ of $SO(10)$, $H_{45}$,
\begin{align}
    \langle H_{45}\rangle =-iv_{45}\times\left(\begin{matrix}
        \sigma_2&0&0&\\
        0&\sigma_2&0&0_{4\times6}&\\
        0&0&\sigma_2&\\
        &0_{6\times 4}&&0_{4\times 4}&\\
    \end{matrix}\right).
    \label{eq:45VEV}
\end{align}
One can see that the bottom-right $4\times4$ block of $H_{45}$ has a vanishing VEV, which we refer to as the ``missing VEV'' of $H_{45}$. As we will see in Sec.~\ref{sec:splitting}, this helps to achieve a mass splitting between dark matter and $SU(3)_c$ colored partners. The VEV in Eq.~\eqref{eq:45VEV} is odd under a discrete subgroup of $SO(10)$ called $C$-parity~\cite{Kibble:1982ae,Lazarides:1985my} (that is also called $D$-parity~\cite{Chang:1983fu,Chang:1984uy}), which involves a charge-conjugation transformation for $SU(3)_c\times U(1)_X$ and the exchange of $SU(2)_L$ with $SU(2)_R$. The VEV is also odd under $CP$. As a result, a linear combination of $C$-parity and $CP$ remains unbroken, which is a left-right symmetry with a space-time parity transformation. We call this symmetry Parity ($P$). The VEV in Eq.~\eqref{eq:45VEV} breaks $SO(10)\times CP$ down to $G_{LR}\times P$.

Is is known that the vacuum in Eq.~\eqref{eq:45VEV} is unstable at tree-level~\cite{Yasue:1980fy,Yasue:1980qj,Anastaze:1983zk} but can be stabilized by quantum corrections via gauge interactions~\cite{Bertolini:2009es}. Alternatively, we may add a $CP$-even Higgs in the $\mathbf{54}$ of $SO(10)$, $H_{54}$, that obtains the following VEV,
\begin{align}
    \langle H_{54}\rangle = \frac{1}{5} v_{54}\times 
\begin{pmatrix}
2 \times \mathbb{1}_{6\times 6} & \\
 & -3 \times \mathbb{1}_{4\times 4}
\end{pmatrix},
    \label{eq:54VEV}
\end{align}
and couples to $H_{45}$~\cite{Babu:1984mz} to stabilize the vacuum at tree-level.

The breaking of $SO(10)\times CP$ into $G_{LR}\times P$ yields massive gauge bosons whose gauge quantum numbers are $({\bf 3},{\bf 2},{\bf 2},1/3)$ and $({\bf 3},{\bf 1},{\bf 1},2/3)$. The former induces proton decay and is called the $XY$ gauge boson. We call the latter the Pati-Salam (PS) gauge boson. The masses of them are 
\begin{align}
\label{eq:MGs}
M_{XY}^2 = g_{10}^2(v_{45}^2+v_{54}^2),~M_{PS}^2 = 4g_{10}^2 v_{45}^2,
\end{align}
where $g_{10}$ is the $SO(10)$ gauge coupling constant.
As we will see, the ratio between these masses,
\begin{equation}\label{eq:rXY_definition}
r_{XY}\equiv\frac{M_{PS}}{M_{XY}},
\end{equation}
affects gauge coupling unification.
When $SO(10)\times CP$ symmetry is broken only by $H_{45}$, $r_{XY}=2$, while a non-zero $v_{54}$ reduces $r_{XY}$.
We will consider $r_{XY}=2$ and $1/2$ as benchmark points.

\subsection{Spontaneous Parity breaking}
\label{sec:ParityBreaking}

We consider the minimal Higgs model, where
$G_{LR}\times P$ is broken down to $G_{\rm SM}$ by the VEV of $H_R$ $(=v_R)$ and $G_{SM}$ is broken down to $SU(3)_c\times U(1)_{\rm EM}$ by the VEV of $H_L$ ($=v_L$). The gauge quantum numbers of $H_R$ and $H_L$ are shown in Table~\ref{tab:SM}. $H_R$ and $H_L$ are Parity partners of each other and are embedded into a ${\bf 16}$ of $SO(10)$, which we call $H_{16}$. Their Parity transformation law is
\begin{align}
H_L(t,{\bf x}) \leftrightarrow H_R^\dag (t,-{\bf x}).
\end{align}

Unlike models with $G_{LR}$ breaking by $SU(2)_R$ triplets and $G_{SM}$ breaking by $SU(2)_L\times SU(2)_R$ bi-fundamentals~\cite{Beg:1978mt,Mohapatra:1978fy}, the Higgs VEVs have no physical phase degree of freedom and a strong $CP$ phase from the phases of the Higgs VEVs is absent. As a result, the strong $CP$ problem can be solved without introducing extra symmetry, as shown in Sec.~\ref{sec:Yukawa_strong_CP}. In models with triplets and bi-fundamentals, the phases of the Higgs VEVs can be suppressed by supersymmetry~\cite{Kuchimanchi:1995rp,Mohapatra:1995xd}, and such models can also be embedded into $SO(10)$~\cite{Mimura:2019yfi}. 

The absence of the $SU(2)_R$ gauge bosons at the electroweak scale requires that $v_R$ be much above $v_L$. Let us discuss how the hierarchy of the VEVs can be obtained through the Higgs Parity mechanism~\cite{Hall:2018let}. The Parity symmetric potential of $H_R$ and $H_L$ at tree-level is
\begin{align}
\label{eq:VH}
V(H_R,H_L)= \lambda\left(|H_R|^2 + |H_L|^2 - f^2\right)^2 + \Delta\lambda |H_R^2||H_L|^2. 
\end{align}
For $\Delta \lambda >0$, the vacua are $(v_L,v_R)=(f,0)$ and $(0,f)$, which are not phenomenologically viable. For $\Delta \lambda <0$, the vacuum is $(v_L,v_R)=(f,f)/\sqrt{2}$, which is also not phenomenologically viable. The only viable possibility is $\Delta \lambda \simeq 0$, for which the vacuum is degenerate at tree-level, $(v_L,v_R)=({\rm cos}\theta,{\rm sin}\theta)f$. The degeneracy is broken by quantum corrections, and we may obtain $v_L\simeq 173$ GeV $\ll v_R \simeq f$ by tuning $\Delta \lambda$ with an accuracy of $v_L^2/v_R^2$ .

This scheme of Parity breaking has phenomenological advantages~\cite{Hall:2018let}. First, despite the existence of the intermediate Parity breaking scale $v_R$, the theory is no more fine-tuned than the SM. The fine-tuning to obtain $v_R$ from a cutoff scale $\Lambda$ is done by the tuning of the parameter $f^2$ with an accuracy of $v_R^2/\Lambda^2$. The fine-tuning to obtain $v_L \ll v_R$ is $v_L^2/v_R^2$. The total degree of fine-tuning is $v_L^2/\Lambda^2$, which is the same as the fine-tuning in the SM with a cutoff scale $\Lambda$. Therefore, if we explain the smallness of $v_L$ by, for example, the anthropic principle~\cite{Agrawal:1997gf,Hall:2014dfa,DAmico:2019hih}, the theory is not fine-tuned beyond what is required from anthropic reasons. This is in contrast to typical $SO(10)$ models with an intermediate scale $v_I$, where the theory has fine-tuning $(v_I^2/\Lambda^2)\times (v_L^2/\Lambda^2) \ll v_L^2 / \Lambda^2$ unless $\Lambda\sim v_I$. 

Second, the Parity breaking scale $v_R$ can be indirectly determined. The potential in Eq.~\eqref{eq:VH} is approximately symmetric under the $SO(2)$ rotation of $H_R$ and $H_L$ when $\Delta\lambda \simeq 0$. The symmetry is spontaneously broken by $\vev{H_R}\neq 0$, and the SM Higgs $H_L$ is understood as a Nambu-Goldstone Boson. In the low energy EFT below $v_R$, the SM Higgs quartic coupling $\lambda_{\rm SM}$ nearly vanishes at the renormalization scale $\sim v_R$, up to a calculable threshold correction. This means that we may determine $v_R$ by precise measurements of SM parameters and computing the renormalization group evolution (RGE) of $\lambda_{\rm SM}$ from the electroweak scale to higher energy scales.

As we will see in Sec.~\ref{sec:gauge_unifcation}, precise $SO(10)$ gauge coupling unification requires a certain range of $v_R$ and predicts a proton decay rate. Therefore, the Higgs Parity mechanism provides a novel connection between precise measurements of SM parameters, gauge coupling unification, and proton decay~\cite{Hall:2019qwx}. In Sec.~\ref{sec:SM}, we show the predictions on SM parameters.

\begin{table}[t]
    \centering
\begin{tabular}{|c||c|c|c|c|c|c|}
 \hline
 $SO(10)$&\multicolumn{6}{|c|}{$\mathbf{16}$}\\
 \hline
 &$q$&\multicolumn{2}{|c|}{$\bar{q}$}&$\ell$, $H_L$&\multicolumn{2}{|c|}{$\bar{\ell}$, $H_R$}\\
 \hline
 $SU(3)$&3&\multicolumn{2}{|c|}{$\bar{3}$}&1&\multicolumn{2}{|c|}{1}\\
 $SU(2)_L$&2&\multicolumn{2}{|c|}{1}&2&\multicolumn{2}{|c|}{1}\\
 $SU(2)_R$&1&\multicolumn{2}{|c|}{2}&1&\multicolumn{2}{|c|}{2}\\
 $U(1)$&1/6&\multicolumn{2}{|c|}{-1/6}&-1/2&\multicolumn{2}{|c|}{1/2}\\
 \hline
 &$q$&$\bar{d}$&$\bar{u}$&$\ell$, $H_L$&$\bar{e}$&$\bar{N}$\\
 \hline
 $SU(3)$&3&$\bar{3}$&$\bar{3}$&1&1&1\\
 $SU(2)_L$&2&1&1&2&1&1\\
 $U(1)$&1/6&1/3&-2/3&-1/2&1&0\\
 \hline
\end{tabular}
\caption{Branching rules of the ${\bf 16}$ of $SO(10)$.}
    \label{tab:SM}
\end{table}

\subsection{Yukawa interactions and the strong $CP$ problem}\label{sec:Yukawa_strong_CP}

Let us now discuss how the SM Yukawa couplings can be obtained and the strong $CP$ problem can be solved. We introduce three ${\bf 16}$ fermions, $\psi_i$, three ${\bf 10}$ fermions, $X_{10,i}$, and three ${\bf 45}$ fermions, $X_{45,i}$. The $G_{\rm LR}$ and $G_{\rm SM}$ decompositions of these fermions are given in Tables~\ref{tab:SM},~\ref{tab:branching_rules_10} and~\ref{tab:branching_rules_45}. The Yukawa interactions of $H_{16}$, $H_{45}$, $\psi_i$ and the $X$-states, and the $SO(10)$ invariant fermion mass terms are
\begin{align}
\label{eq:L10}
{\cal L} = & - x_{10}^{ij}H_{16} \psi_i X_{10,j} - i x_{10}^{'ij} H_{16} \psi_i X_{10,j} H_{45} - (M_{10}^{ij} + i\lambda_{10}^{ij}H_{45}) X_{10,i}X_{10,j} \nonumber \\
&- x_{45}^{ij}H_{16}^\dag \psi_i X_{45,j} - i x_{45}^{'ij}H_{16}^\dag \psi_i X_{45,j} H_{45} - (M_{45}^{ij} + i\lambda_{45}^{ij}H_{45}) X_{45,i}X_{45,j} + {\rm h.c.},
\end{align}
where all the parameters are real due to $CP$ symmetry, $M_{ij}$ is symmetric, and $\lambda_{ij}$ is anti-symmetric. The theta term of the $SO(10)$ gauge field is $0$ or $\pi$.

The VEV of $H_{45}$ gives complex phases to the Yukawa interactions and masses. However, the residual Parity symmetry guarantees that the strong $CP$ phase remains zero. For example, the down-type Yukawa couplings come from the Dirac masses and Yukawa interactions of $q$, $\bar{q}$, $D$, and $\bar{D}$, i.e., the masses and couplings of $X_{10,i}$ in the first line of Eq.~\eqref{eq:L10}. Their Parity transformation law is
\begin{align}
    q(t,{\bf x}) \leftrightarrow i \sigma_2 \bar{q}^*(t,-{\bf x}),~~D(t,{\bf x}) \leftrightarrow i \sigma_2 \bar{D}^*(t,-{\bf x}).
\end{align}
The Parity-invariant masses and Yukawa interactions of them are
\begin{align}
    {\cal L} = - x_d^{ij}H_L q_i \bar{D}_j -  x_d^{*ij}H_R \bar{q}_i D_j - M_d^{ij}D_i \bar{D}_j + {\rm h.c.}, 
\end{align}
where $M_d$ is Hermitian. The real/complex parts of $x_d$ and $M_d$ come from the $H_{45}$ independent/dependent terms in Eq.~\eqref{eq:L10}. The mass matrix of $d \subset q$, $\bar{d} \subset \bar{q}$, $D$, and $\bar{D}$ is
\begin{align}
    \begin{pmatrix}
    d_i & D_i
    \end{pmatrix}
    \begin{pmatrix}
    0 & x_d^{ij} v_L  \\
    x_d^{*ji} v_R & M_d^{ij}
    \end{pmatrix}
    \begin{pmatrix}
    \bar{d}_j \\ \bar{D}_j
    \end{pmatrix}.
\end{align}
The determinant of the mass matrix is real and the contribution to the strong $CP$ phase from down-type quarks is $0$ or $\pi$ at leading order. Similarly, the up sector contributes to the strong $CP$ phase by $0$ or $\pi$. As a result, the strong $CP$ phase is $0$ or $\pi$ at leading order, and for the former case, the strong $CP$ problem is solved~\cite{Babu:1988mw,Babu:1989rb,Hall:2018let}. A non-zero strong $CP$ phase arises at loop level, but the correction can be below the experimental upper bound~\cite{Hall:2018let,deVries:2021pzl,Hisano:2023izx}. 

The down-type Yukawa is determined in the following way. If $M_d \gg x_d v_R$, we may integrate out $D$ and $\bar{D}$ to obtain an effective interaction $q  x_d M_d^{-1} x_d^\dag \bar{q} H_L H_R $. The SM right-handed down quark is $\bar{d}$, and the down-type Yukawa is given by $x_d M_d^{-1} x_d^\dag v_R$. If $M_d \ll x_d v_R$, $\bar{d}$ becomes a Dirac partner of $D$ with mass $x_d v_R$. The SM right-handed down quark is $\bar{D}$ with a Yukawa coupling $x_d$. The up-type Yukawa is determined in a similar way from the masses and Yukawa interactions of $X_{45,i}$ in the second line of Eq.~\eqref{eq:L10}. See \cite{Hall:2019qwx} for details.

The electron-type Yukawa couplings also come from the masses and couplings of $X_{10,i}$ in the first line of Eq.~\eqref{eq:L10},
\begin{align}
    {\cal L} = - x_e^{ij}H_R \ell_i \Delta_j -  x_e^{*ij}H_L \bar{\ell}_i \Delta_j - \frac{1}{2}M_e^{ij}\Delta_i \Delta_j + {\rm h.c.}, 
\end{align}
where $M_e$ is real and symmetric. Because  of $x_e \neq x_d$ and $M_e\neq M_d$, arising from the VEV of $H_{45}$, $m_e\neq m_d$ can be explained. See \cite{Hall:2019qwx} for the discussion of neutrino masses and mixing.

\section{Electroweak-Charged Dark Matter}\label{sec:WIMP_DM}

In this section, we discuss electroweak-charged dark matter in $SO(10)$. We focus on dark matter that is embedded into a ${\bf 10}$ of $SO(10)$. We will comment on ${\bf 45}$ and ${\bf 54}$, which are subject to stronger constraints and have no viable parameter space, at the end of Sec.~\ref{sec:gauge_unifcation}.

\subsection{Dark matter phenomenology }\label{sec:WIMPDM}

We assume that dark matter is fermionic to avoid an extra fine-tuning problem related to the mass scale of dark matter. To stabilize the dark matter, we impose $\mathbb{Z}_2$ symmetry on a Weyl fermion embedded in a ${\bf 10}$ of $SO(10)$, which we call $\chi_{10}$, that branches to $({\bf 1},{\bf 2},-1/2) \equiv \chi_L$, $({\bf 1},{\bf 2},1/2) \equiv \chi_{\bar{L}}$, $({\bf 3},{\bf 1},-1/3) \equiv \chi_D$, and $({\bf \bar{3}},{\bf 1},1/3) \equiv \chi_{\bar{D}}$ of $G_{SM}$, with the following Dirac masses,
\begin{align}
    {\cal L} = - m_{L} \chi_{\bar{L}} \chi_{L}  - m_{D} \chi_{\bar{D}} \chi_{D} + {\rm h.c.}.
\end{align}
Hereafter, we will call the Dirac states $\chi_L$ and $\chi_D$. 
We assume $m_{D} \gg m_{L}$, which can be achieved by a coupling of $\chi_{10}$ with $H_{45}$, as we will see in Sec.~\ref{sec:splitting}. If the reheating temperature of the universe $T_R$ is much below $m_{D}$, then $\chi_D$ is not produced in the early universe. Even if $\chi_D$ is produced, it can decay into $\chi_L$ early enough without causing cosmological problems so long as the mass splitting is large enough.

In both cases, only $\chi_L$ has a non-negligible abundance in the present universe and may explain the observed dark matter density. The dark matter phenomenology of $\chi_L$ is essentially the same as Higgsino-like dark matter in supersymmetric theories. Assuming $T_R > m_{L}/20$, the freeze-out mechanism explains the observed dark matter abundance if $m_{L}\simeq 1$ TeV~\cite{Olive:1990qm,Cirelli:2005uq}. As we will see later, the decay of $\chi_D$ can explain the observed dark matter density even if $m_{L} < 1$ TeV. In this case, $m_{L}$ may be as small as the LEP bound of $100$ GeV~\cite{LEP}.

If the $\chi_L$ do not mix with other states and remain Dirac particles, they interact with nucleons via $Z$-boson exchange without suppression by the velocity of dark matter, which is excluded by direct-detection experiments. To be a viable dark matter candidate, they should mix with other states to become Majorana particles. The simplest possibility would be mixing with an electroweak singlet $S$ with a Majorana mass. At the $SO(10)$ level, the interaction and mass terms of $S$ are
\begin{align}
    {\cal L} = - \frac{1}{2M} S \chi_{10} H_{16} H_{16} - \frac{1}{2M'} S \chi_{10} H_{16}^\dag H_{16}^\dag - \frac{1}{2}m_S S^2 + {\rm h.c.}.
\end{align}
The first two terms can be UV-completed by, e.g, the exchange of fermions embedded in a ${\bf 16}$ of $SO(10)$. For simplicity we assume $M' \gg M$ and $m_S \gg m_{L}$. Then, after taking $\vev{H_R}=v_R$ and integrating out $S$, we obtain
\begin{align}
   {\cal L} = \frac{v_R^2}{2M^2 m_S} \chi_L \chi_L H_L H_L. 
\end{align}
After electroweak symmetry breaking, the neutral component of $\chi_L$ obtains a Majorana mass term. Then the neutral components of $\chi_L$ and $\chi_{\bar{L}}$ split into two Majorana fermions $\chi_1$ and $\chi_2$ with mass splitting
\begin{align}
\Delta m_0 = m_{\chi_2}- m_{\chi_1} \simeq \frac{v_R^2 v_L^2}{M^2 m_S} = 100~{\rm keV}~\frac{10~{\rm TeV}}{m_S} \left(\frac{v_R/M}{0.006}\right)^2.
\end{align}
As long as $\Delta m_0 \gtrsim 100 $ keV, the up scattering in direct-detection experiments $\chi_1 N \rightarrow \chi_2 N$, where $N$ is a nucleon, is kinematically forbidden. This requires that the mass scale $M$ is not much above $v_R$. The scattering $\chi_1 N \rightarrow \chi_1 N$ is suppressed by the velocity of dark matter and does not constrain the model.

Dark matter can also be probed by indirect-detection experiments. If the dark matter halo profile at the center of the galaxy is cuspy enough, gamma-ray observations can detect the annihilation of dark matter~\cite{Rinchiuso:2020skh}.

The collider search for dark matter typically relies on disappearing tracks from the decay of the charged component into the neutral component of $\chi_L$, and depends on the mass difference $\Delta m_\pm $ between them. If dominated by electroweak quantum corrections, $\Delta m_\pm \simeq 340$~MeV, and the bound on $m_L$ is the LHC bound of 200 GeV~\cite{ATLAS:2022rme,CMS:2023mny}. High-Luminosity LHC can probe the dark matter mass up to $500$ GeV~\cite{Fukuda:2017jmk}. If $\Delta m_0$ becomes comparable to the electroweak correction, $\Delta m_\pm$ becomes larger and the collider bound on $m_L$ weakens down to the LEP bound of 100 GeV.

\subsection{Cosmological constraints on colored partners}\label{sec:decay_of_colour_particles}

Let us now discuss the constraint on $m_{D}$, first assuming $T_R > m_{D}/20$. $\chi_D$ is abundantly produced in the early universe via $SU(3)_c$ interactions. As the temperature drops below $m_{D}$, the abundance of $\chi_D$ is exponentially suppressed. The annihilation of them freezes-out at around $T\sim m_D/20$, and the resultant number density of $\chi_D$ is
\begin{align}
\label{eq:chidnumber}
\frac{n_{\chi_D}}{s} \simeq 10^{-7} \times \left( \frac{m_{D}}{10^{10}~{\rm GeV}}\right)^2.
\end{align}
If $\chi_D$ decays after the QCD phase transition, the number density of $\chi_D$ decreases further before decay~\cite{DeLuca:2018mzn}. Around the QCD phase transition, $\chi_D$ forms bound states with SM quarks. The bound states have large radii $\sim \Lambda_{\rm QCD}^{-1}$ and scatter with each other efficiently. The scattering produces bound states made from $\chi_D$ and its anti-particle, which decay into gluons. As a result, the number density of bound states made of $\chi_D$ and SM quarks decreases exponentially. However, the scattering also produces bound states made from three $\chi_D$, which are stable up to the decay into $\chi_L$ via $XY$ gauge boson exchange. The number density of such bound states is of the same order as the original $\chi_D$ density in Eq.~\eqref{eq:chidnumber}. 

Depending on when $\chi_D$ decays, there are constraints from the overproduction of dark matter. $\chi_D$ decays into dark matter and SM particles via $XY$ gauge boson exchange. The decay rate is 
\begin{align}
\Gamma \sim  \frac{1}{128\pi^3} \frac{m_{D}^5}{m_{XY}^4},
\end{align}
and the decay occurs at around the temperature
\begin{align}
T_{\rm dec} \simeq&~2~{\rm MeV} \left(\frac{m_{D}}{10^9~{\rm GeV}}\right)^{5/2} \left(\frac{10^{16}~{\rm GeV}}{m_{XY}}\right)^2.
\end{align}

When the decay of $\chi_D$ occurs before the freeze-out of $\chi_L$ annihilation at around $T_{\rm FO}\simeq m_{L}/20$, the $\chi_L$ produced via the decay are thermalized and the dark matter abundance is determined by the freeze-out of $\chi_L$ annihilation. Even when the decay occurs after the freeze-out of $\chi_L$ annihilation, the number density can decrease by annihilation down to a density of $n \simeq H/(\sigma v)$. The resultant dark matter density is
\begin{align}
   \frac{\rho_{\rm DM}}{s} \simeq 0.4~{\rm eV} \left(\frac{m_{L}}{100~{\rm GeV}}\right)^3 \frac{0.05~{\rm GeV}}{T_{\rm dec}}.
\end{align}
The coefficient of this formula is determined so that the observed dark matter density $\rho_{\rm DM} \simeq 0.4$ eV is reproduced when $m_{L}=1$ TeV and $T_{\rm decay}= T_{\rm FO} \simeq 50$ GeV. To avoid the overproduction of dark matter, it is required that
\begin{align}
\label{eq:mDlow}
m_{D} > 3 \times 10^{9}~{\rm GeV}\times \left(\frac{m_{XY}}{10^{16}{\rm GeV}}\right)^{4/5} \left(\frac{m_{L}}{100{\rm GeV}}\right)^{6/5}.
\end{align}
When $m_{L} < 1$ TeV, this bound should be saturated so that the observed dark matter abundance can be explained by the production of dark matter via the decay of the colored partner. Note that the bound from dark matter overproduction requires that $\chi_D$ decays much before BBN begins, so that the BBN bound is satisfied as long as the overproduction bound is satisfied.

We next discuss the possibility of $T_R < m_D/20$. In this case, the abundance of $\chi_D$ can be suppressed in comparison to the case with $T_R > m_D/20$, and by taking $T_R \ll m_D/20$, the cosmological bound in Eq.~\eqref{eq:mDlow} can be avoided. See~\cite{Chung:1998rq,Kurata:2012nf,Harigaya:2014waa,Harigaya:2016vda,Harigaya:2019tzu} for the estimation of the abundance. In the most conservative case, assuming that the maximal temperature of the universe $T_{\rm max}$ is as large as $T_R \sim m_L/20$, even if $m_D$ is only a factor of a few larger than $m_L$, the abundance of $\chi_D$ is exponentially suppressed in comparison to that of $\chi_L$, and the cosmological bound on $m_D$ and $M_{XY}$ can be avoided. $T_{\rm max}\simeq T_R$ is, in principle, possible in certain reheating scenarios~\cite{Felder:1998vq,Co:2020xaf}. 

For low $T_R$, however, the most economical way to generate baryon asymmetry-- leptogenesis-- becomes difficult. Parity predicts right-handed neutrinos whose out-of-equilibrium decay can generate lepton asymmetry. Unless right-handed neutrinos are non-thermally produced by the decay of an inflaton and/or are degenerate in their masses, successful leptogenesis requires $T_R>2\times 10^9$ GeV~\cite{Giudice:2003jh,Buchmuller:2004nz}. Then violating the assumption of $T_R>m_D/20$ requires
\begin{equation}
\label{eq:mDlowweak}
    m_D > 4 \times 10^{10}~{\rm GeV}.
\end{equation}
As we will see, this bound is still strong and the parameter space cannot be expanded much.%
\footnote{
Furthermore, unless $T_{\rm max}\sim T_R$, non-negligible amounts of $\chi_D$ are still produced before the completion of reheating and thermalization~\cite{Chung:1998rq,Kurata:2012nf,Harigaya:2014waa,Harigaya:2016vda,Harigaya:2019tzu}, and the lower bound on $m_D$ can be even stronger. 
}

In Sec.~\ref{sec:cosmo_proton_constraints}, we discuss the implications of the bounds in Eqs.~\eqref{eq:mDlow} and \eqref{eq:mDlowweak} to the proton decay rate and the prediction on $v_R$.

\subsection{Mass splitting}
\label{sec:splitting}
In this section, we show how to obtain a mass splitting between the dark matter particles $\chi_L$ and the colored partners $\chi_D$. We first compute the mass splitting for a single $\chi_{10}$ Weyl fermion in $SO(10)$. We will find that although we can obtain $m_D\gg m_L$ at tree-level, 1-loop mass corrections generate $m_L$ and destabilize the mass splitting. To fix this issue, we will consider two $\mathbf{10}$ Weyl fermions in $SO(10)$, for which we find sufficiently small quantum corrections to $m_L$.

\subsubsection{One Weyl fermion}
Consider one $\chi_{10}$ Weyl fermion. The first six components of $\chi_{10}^a$ $(a=1,2,\cdots,10)$ contain $\chi_D$ and $\chi_{\bar{D}}$, and the last four components contain $\chi_L$ and $\chi_{\bar{L}}$.

In order to achieve a mass splitting with $m_D\gg m_L$, we couple $\chi_{10}$ to $H_{45}$. Note that the term $\chi_{10} H_{45}\chi_{10}$ identically vanishes because of the anti-symmetric $SO(10)$ indices of $H_{45}$ and the Fermi statistics of $\chi_{10}$. We thus consider a higher order term,
\begin{equation}\label{eq:1genH45coupling}
    \chi_{10}^a H_{45}^{ab} H_{45}^{bc}\chi_{10}^c.
\end{equation}
Because of the missing VEV of $H_{45}$, this interaction gives a mass only to $\chi_D$ and gives a large mass splitting between $\chi_D$ and $\chi_L$. This mechanism is analogous to the missing VEV mechanism for the doublet-triplet splitting~\cite{Dimopoulos:1981xm}.

If $H_{45}$ is a real field, however, the mass splitting is quantum mechanically unstable. This is because quadratically divergent quantum corrections generate a quadratic term $\chi_{10}^a \chi_{10}^a$, which gives the same mass term to $\chi_D$ and $\chi_L$. If $H_{45}$ is a complex field, the quadratically divergent correction is absent, but still a term with different $SO(10)$ index contraction, $\chi_{10}^a \chi_{10}^a H_{45}^{bc} H_{45}^{bc}$, is generated by quantum corrections via $SO(10)$ gauge interactions. The natural mass splitting is at most $g^2/(16\pi^2)\sim 10^{-3}$, which is not large enough to satisfy the cosmological bounds in Eqs.~\eqref{eq:mDlow} or \eqref{eq:mDlowweak}.

\subsubsection{Two Weyl fermions}

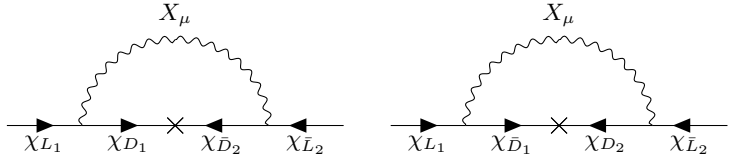
\begin{figure*}[t]
    \centering
     \begin{subfigure}[b]{0.32\textwidth}
         \centering
         \begin{tikzpicture}[baseline=(current bounding box.center)]
  \begin{feynman}
    \vertex (x);
    \vertex[right=0.5\textwidth of x] (y) ;
    \vertex[right=0.25\textwidth of x] (z);
    \vertex[left=0.2\textwidth of x] (a);
    \vertex[right=0.2\textwidth of y] (d);
    \vertex[above=0.23\textwidth of z] (w);
    \vertex[above=0.01\textwidth of w] (f1) {\footnotesize\(X_{\mu}\)};

    \diagram*{
        (x) --[boson, quarter left] (w),
        (w) --[boson, quarter left] (y),
        (x) --[fermion,insertion=0.9999, edge label'=\footnotesize\(\chi_{D_1}\)] (z),
        (z) --[anti fermion, edge label'=\footnotesize\(\chi_{\bar{D}_2}\)] (y),
        (a) --[fermion, edge label'=\footnotesize\(\chi_{L_1}\)] (x),
        (y) --[anti fermion, edge label'=\footnotesize\(\chi_{\bar{L}_2}\)] (d),
    };
  \end{feynman}
\end{tikzpicture}
         \label{fig:1genScalarB}
     \end{subfigure}
    \begin{subfigure}[b]{0.32\textwidth}
         \centering
         \begin{tikzpicture}[baseline=(current bounding box.center)]
  \begin{feynman}
    \vertex (x);
    \vertex[right=0.5\textwidth of x] (y) ;
    \vertex[right=0.25\textwidth of x] (z);
    \vertex[left=0.2\textwidth of x] (a);
    \vertex[right=0.2\textwidth of y] (d);
    \vertex[above=0.23\textwidth of z] (w);
    \vertex[above=0.01\textwidth of w] (f1) {\footnotesize\(X_{\mu}\)};

    \diagram*{
        (x) --[boson, quarter left] (w),
        (w) --[boson, quarter left] (y),
        (x) --[fermion,insertion=0.9999, edge label'=\footnotesize\(\chi_{\bar{D}_1}\)] (z),
        (z) --[anti fermion, edge label'=\footnotesize\(\chi_{D_2}\)] (y),
        (a) --[fermion, edge label'=\footnotesize\(\chi_{L_1}\)] (x),
        (y) --[anti fermion, edge label'=\footnotesize\(\chi_{\bar{L}_2}\)] (d),
    };
  \end{feynman}
\end{tikzpicture}
         \label{fig:intermediate}
     \end{subfigure}

    \caption{Possible 1-loop corrections to the mass of $\chi_L$ via gauge interactions.}
    \label{fig:1genDiagrams}
\end{figure*}

Large mass splitting between $\chi_D$ and $\chi_L$ is possible if there are two $\mathbf{10}$ Weyl fermions, $\chi_{10_1}$ and $\chi_{10_2}$, and a Yukawa interaction
\begin{equation}\label{eq:2genH45coupling}
    i y \chi_{10_1}^a H_{45}^{ab} \chi_{10_2}^b + {\rm h.c.}.
\end{equation}
This interaction gives the same mass to $\chi_{D_1} \chi_{\bar{D}_2}$ and $\chi_{D_2} \chi_{\bar{D}_1}$, and does not give mass to $\chi_L$ at tree-level. Quantum corrections do not generate mass terms for $\chi_L$ for the following reason. The interaction in Eq.~\eqref{eq:2genH45coupling} preserves a $\mathbb{Z}_4$ symmetry under which $\chi_{10_{1,2}}$ has charge $1$ and $H_{45}$ has charge $2$, so any mass terms of $\chi_{10_{1,2}}$ generated by quantum corrections involve an odd number of $H_{45}$. Then to obtain a non-zero mass, the $SO(10)$ indices of $\chi_{10_{1,2}}$ must be contracted with $H_{45}$, and only $\chi_{D}$ obtains a non-zero mass. One can explicitly confirm the absence of corrections. For example, 1-loop corrections via gauge interactions are given by the diagrams in Fig.~\ref{fig:1genDiagrams}. The corrections from the two diagrams cancel with each other because of the opposite signs of the masses of $\chi_{D_1} \chi_{\bar{D}_2}$ and $\chi_{D_2} \chi_{\bar{D}_1}$.

To give a non-zero mass to $\chi_L$, we may add $\chi_{10_1} \chi_{10_2}$, which gives the same mass to $\chi_{L_1} \chi_{\bar{L}_2}$ and $\chi_{L_2} \chi_{\bar{L}_1}$. In this case, however, two pairs of dark matter fermions affect the gauge coupling unification through the RGE running from the dark matter mass scale to the colored particle mass scale, which lowers the unification scale so much that the proton decays too rapidly. To avoid this, we instead add $m_2\chi_{10_2}\chi_{10_2}$. By taking $m_2 \gtrsim y v_{45}$, only one pair of dark matter fermions affect the gauge coupling unification. Among the two mass eigenstates of $\chi_L/\chi_D$s, we call the lighter $\chi_{L}/\chi_D$ and the heavier $\chi_{L'}/\chi_{D'}$. Note that $m_{D'} \geq m_{L'}$ in the setup described above. 

The quantum corrections by $H_{45}$ generate a mass term of $\chi_{10_1}\chi_{10_1}$ as large as $y^2 m_2/(16\pi^2)$. For example, for $v_{45}\sim 10^{16}$ GeV, $y\sim 10^{-4}$, and $m_2 \sim 10^{12}$ GeV, the quantum corrections to the mass are as small as $100$ GeV and do not disturb the assumed mass splitting.

The missing VEV of $H_{45}$ is stabilized by the $SU(2)_R$ symmetry under which the bottom-right component in Eq.~\eqref{eq:45VEV} is charged. Once $SU(2)_R$ symmetry is broken, there is no symmetry preventing a VEV of the bottom-right component. Indeed, the following coupling,
\begin{align}
\lambda_{16,45}H_{16}\Gamma^{abcd} H_{16} H_{45}^{ab} H_{45}^{cd}
\end{align}
gives a tadpole term of the bottom-right component with a coefficient $\sim \lambda_{16,45} v_R^2 v_{45}$. Then the bottom-right component obtains a VEV $\sim v_{45} (v_R/v_{\rm 45})^2 (\lambda_{16,45}/\lambda_{45})$, where $\lambda_{45}$ is the quartic coupling of $H_{45}$. When $SO(10)\times CP$ is broken to $G_{LR}\times P$ only by $H_{45}$, the vacuum is unstable at tree-level and is stabilized by quantum corrections. This requires $\lambda_{45}\sim \alpha_{10}^2\sim 10^{-3}$, where $\alpha_{10}$ is the fine-structure constant of $SO(10)$. To be conservative, we take $\lambda_{16,45}\sim 10^{-3}$, which is as small as is generated by quantum corrections via $SO(10)$ gauge interactions. In the viable parameter space with precise gauge coupling unification that we identify in Sec.~\ref{sec:cosmo_proton_constraints}, $v_R<$~few$ \times 10^{11}$ GeV and $v_{45} > 10^{16}$ GeV, for which $m_D/m_L < 10^9$ is stable against the correction to $m_L$ by the tadpole term of the bottom-right component of $H_{45}$. The mass splitting remains large enough to satisfy the cosmological constraints.

\section{Gauge Coupling Unification}\label{sec:gauge_unifcation}
In this section, we discuss the running of the gauge couplings, the quality of unification, and the constraints from cosmological and proton decay bounds.

\subsection{Gauge coupling running}
\label{sec:gauge running}
We perform 2-loop RGE on the gauge couplings from $m_Z$ to the unification scale $M_{XY}$. The RGE is solved in two regimes between $m_Z$ and $M_{XY}$: from $m_Z$ to $M_{W_R}$ with gauge group $G_\text{SM}$ and from $M_{W_R}$ to $M_{XY}$ with gauge group $G_\text{LR}$, where $M_{W_R}$ is the mass of the heavy $SU(2)_R$ gauge boson. Due to the left-right symmetry of $G_\text{LR}$, the running of the $SU(2)_L$ and $SU(2)_R$ gauge couplings are identical, and we will refer to both gauge couplings as $g_2$. The $U(1)$ gauge coupling will be written in $SO(10)$ normalization for both $G_{\text{SM}}$ and $G_{\text{LR}}$, and in both cases we will refer to the gauge coupling as $g_1$. Superscripts SM and LR will be used if there is ambiguity.

The 2-loop RGE of fine-structure constant $\alpha_i\equiv g_i^2/(4\pi)$ is given by
\begin{equation}
    \dv{}{\ln\mu}\left(\frac{2\pi}{\alpha_i}\right)=b_i+\sum_j b_{ij}\frac{\alpha_j}{2\pi},
\end{equation}
where
\begin{equation}
    b_i=\left(\begin{matrix}
        b_1\\b_2\\b_3 \end{matrix}\right),~b_{ij}=\left(\begin{matrix}
        b_{11}&b_{12}&b_{13}\\b_{21}&b_{22}&b_{23}\\b_{31}&b_{32}&b_{33}
    \end{matrix}\right)
\end{equation}
are the 1-loop and 2-loop $\beta$-function coefficients, respectively, and $i,j=1,2,3$. At renormalization scale $\mu$, the RGE is solved with
\begin{equation}
    b_i=\sum_{n,\,m_n\leq \mu}b_i^{n},~b_{ij}=\sum_{n,\,m_n\leq \mu}b_{ij}^{n},
\end{equation}
where the sum is over particles, labelled by $n$, with mass equal to or below $\mu$.

For $G_{\text{SM}}$ with only SM particle content, $b_i$ and $b_{ij}$ are
\begin{equation}\label{eq:SM_beta_coeffs}
b^{\text{SM}}_i=\left(\begin{matrix}
        -41/10\\19/6\\7
    \end{matrix}\right),~b^{\text{SM}}_{ij}=\left(\begin{matrix}
        -\frac{199}{100}&-\frac{27}{20}&-\frac{22}{5}\\-\frac{9}{20}&-\frac{35}{12}&-6\\-\frac{11}{20}&-\frac{9}{4}&13
    \end{matrix}\right),
\end{equation}
and for $G_{\text{LR}}$ with only SM particle content and their Parity partners (i.e., right-handed neutrinos, $H_R$, and $SU(2)_R\times U(1)_{X}$ gauge bosons),
\begin{equation}  
    b^{\text{LR}}_i=\left(\begin{matrix}
        -9/2\\19/6\\7
\end{matrix}\right),~b^{\text{LR}}_{ij}=\left(\begin{matrix}
        -\frac{23}{8}&-\frac{27}{4}&-2\\-\frac{9}{8}&-\frac{35}{12}&-6\\-\frac{1}{4}&-\frac{9}{2}&13
    \end{matrix}\right).
\end{equation}

In addition to SM particles and their Parity partners, we include contributions from the dark matter multiplets $\chi_{10_{1,2}}$ that branch to $\chi_L,\chi_{L'},\chi_D,\chi_{D'}$ in $G_{\text{SM}}$; see Sec.~\ref{sec:WIMP_DM}. The contributions of $\chi_{10_{1,2}}$ to $b_i$ and $b_{ij}$ for both $G_{\text{SM}}$ and $G_{\text{LR}}$ are shown in Appendix~\ref{app:BR_beta_fn}. We take the dark matter mass $m_L$ to be between $200\textrm{ GeV}$ and $1\textrm{ TeV}$. The colored partner mass $m_{D}$ is taken to be between $m_{L}$ and $10^{14}\textrm{ GeV}$. As we will see in Sec.~\ref{sec:cosmo_proton_constraints}, cosmological and proton decay bounds will constrain the allowed values of $m_{D}$. We take $m_{L'}=m_{D'}=10^{12}~\textrm{GeV}$, which allows for sufficient mass splitting between $\chi_L$ and $\chi_D$; see Sec.~\ref{sec:splitting}. 

We also include the six $X$-states, $X_{10,i}$ and $X_{45,i}$, that are required to produce the SM Yukawa couplings; see Sec.~\ref{sec:Yukawa_strong_CP}. The contributions of the $X$-states to $b_i$ and $b_{ij}$ for both $G_{\text{SM}}$ and $G_{\text{LR}}$ are shown in Appendix~\ref{app:BR_beta_fn}. We determine the masses of the $X$-states in the following way. With the $G_\text{SM}$ gauge couplings, we compute the RGE of the SM Yukawa couplings from $m_Z$ to $M_{W_R}$ and use the values of the up-type and down-type Yukawa couplings at $M_{W_R}$ to determine the masses of the six $X$-states, taking $x=1$. We take the masses of the $X$-state particles in the same $SO(10)$ multiplets to be universal. There can be $\mathcal{O}(1)$ mass splitting between colored and non-colored particles in the $X$-states. We comment on their effects later.

We solve the RGE equations from $m_Z$ to $M_{W_R}$, modifying the RGE $\beta$-function coefficients as described above, and matching the gauge couplings to experimental values in the $\overline{\text{MS}}$ scheme at the renormalization scale $m_t$~\cite{Buttazzo:2013uya},
\begin{equation}
    g_1(m_t)=0.4626,~g_2(m_t)=0.64779,~g_3(m_t)=1.1666.
\end{equation}
The $G_{SM}$ gauge couplings are matched to the $G_{\text{LR}}$ gauge couplings via the 1-loop matching conditions
\begin{equation}
\begin{split}
     &\frac{2\pi}{\alpha^{SM}_1(M_{W_R})}=\frac{2}{5}\frac{2\pi}{\alpha^{LR}_{1}(M_{W_R})}+\frac{3}{5}\frac{2\pi}{\alpha_2^{LR}(M_{W_R})}-\frac{1}{10},\\
     &\frac{2\pi}{\alpha^{SM}_2(M_{W_R})}=\frac{2\pi}{\alpha^{LR}_2(M_{W_R})},~\frac{2\pi}{\alpha^{SM}_3(M_{W_R})}=\frac{2\pi}{\alpha^{LR}_3(M_{W_R})}.
\end{split}
\end{equation}
After the RGE above $M_{W_R}$, the $G_\text{LR}$ gauge couplings are matched to the the $SO(10)$ gauge coupling at the mass $M_{XY}$ of the $XY$ gauge boson of charge $({\bf 3},{\bf 2},{\bf 2},-1/3)$ (see Sec.~\ref{sec:SO(10)_breaking}),
\begin{equation}\label{eq:SO(10)_matching}
    \begin{split}
        \frac{2\pi}{\alpha_1(M_{XY})}&=\frac{2\pi}{\alpha_{10}(M_{XY})}+\Delta_{1,G}+\Delta_{1,H}+\Delta_1,\\
        \frac{2\pi}{\alpha_2(M_{XY})}&=\frac{2\pi}{\alpha_{10}(M_{XY})}+\Delta_{2,G} +\Delta_{2,H}+\Delta_2,\\
        \frac{2\pi}{\alpha_3(M_{XY})}&=\frac{2\pi}{\alpha_{10}(M_{XY})}+\Delta_{3,G} +\Delta_{3,H}+\Delta_3,
    \end{split}
\end{equation}
where $\Delta_{i,G}$ are threshold corrections from heavy gauge bosons, $\Delta_{i,H}$ are threshold corrections from $SO(10)\times CP$ breaking Higgses, and $\Delta_{i}$ are possible extra threshold corrections. The corrections to the gauge couplings from the possible mass splitting of $X$-states can be included in $\Delta_i$. For a given unification scale $M_{XY}$, the required threshold corrections beyond those from heavy gauge and Higgs bosons can be parameterized by
\begin{equation}
    \Delta(M_{XY})\equiv\max_{i,j}\abs{\Delta_i-\Delta_j}=\max_{i,j}\abs{\left(\frac{2\pi}{\alpha_i}-\Delta_{i,G}-\Delta_{i,H}\right)-\left(\frac{2\pi}{\alpha_j}-\Delta_{j,G}-\Delta_{j,H}\right)}.
\end{equation}

The threshold corrections from heavy gauge bosons are
\begin{equation}\label{eq:heavy_gauge_bosons_thresh}
    \begin{split}
        \Delta_{1,G}=14\ln r_{XY}-\frac{4}{3},~\Delta_{2,G}=-1,~\Delta_{3,G}=\frac{7}{2}\ln r_{XY}-\frac{5}{6},
    \end{split}
\end{equation}
where $r_{XY}$ is the ratio between the gauge boson of charge $({\bf 3},{\bf 2},{\bf 2},-1/3)$ and the gauge boson of charge $({\bf 3},{\bf 1},{\bf 1},2/3)$; see Eq.~\eqref{eq:rXY_definition}. We consider benchmark values $r_{XY}=2$ and $1/2$. The threshold corrections from the $H_{45}$ Higgs field are
\begin{equation}
    \Delta_{1,H}=0,~\Delta_{2,H}=-\frac{1}{3}\ln\frac{M_{(1,3,1,0)}}{M_{XY}}=-\frac{1}{3}\ln\frac{M_{(1,1,3,0)}}{M_{XY}},~\Delta_{3,H}=-\frac{1}{2}\ln\frac{M_{(8,1,1,0)}}{M_{XY}},
\end{equation}
where $M_{(8,1,1,0)}$, $M_{(1,3,1,0)}$ and $M_{(1,1,3,0)}$ are the masses of the physical Higgs fields after $SO(10)\times CP$ breaking with subscripts denoting their $G_\text{LR}$ charges. When $SO(10)\times CP$ is broken solely by $H_{45}$, the vacuum is unstable at tree-level and is stabilised by quantum corrections via gauge interactions. This requires that the quartic coupling of $H_{45}$ is $\mathcal{O}(\alpha_{10}^2)$. To be concrete, we take the tree-level quartic to be zero, for which~\cite{Bertolini:2009es}
\begin{equation}
\label{eq:Hmass}
    \frac{M_{1,3,1,0}^2}{M_{XY}^2}=\frac{M_{1,1,3,0}^2}{M_{XY}^2} =\frac{19g^2}{4\pi^2},~\frac{M_{8,1,1,0}^2}{M_{XY}^2} =\frac{22g^2}{4\pi^2}.
\end{equation}
If the tree-level quartic is non-zero, the physical Higgs masses can be different from Eq.~\eqref{eq:Hmass} by $\mathcal{O}(1)$ factors, which can be taken into account by $\Delta_i$ of $\mathcal{O}(0.1)$. If $v_{54}$ is comparable to $v_{45}$, the physical Higgs masses can be comparable to $M_{XY}$, but that can also be taken into account by $\Delta_i$ of $\mathcal{O}(1)$. 

In the minimal setup we consider, $\Delta(M_{XY})$ is expected to be $\mathcal{O}(1)$. There can be threshold corrections from the mass splitting of $X$-states via their couplings with $SO(10)\times CP$ breaking Higgses. The mass splitting of $X_{10,j}$ cannot be more than $\mathcal{O}(1)$, since otherwise the electron-type and down-type Yukawa couplings are split too much. The contribution to $\Delta$ from $X_{10,j}$ is therefore at most $\mathcal{O}(1)$.%
\footnote{
Also, $m_{e}/m_d \simeq m_{s}/m_\mu\simeq 1/3$ around the unification scale. If the mass difference is explained by the mass splitting of $X_{10,j}$, the contribution to $\Delta_i$ from $X_{10,1}$ approximately cancels with that from $X_{10,2}$.
}
The mass splitting of $X_{45,j}$ can be larger. In particular, couplings with $H_{45}$ can naturally make colored particles much heavier than non-colored particles, for which $X_{45,j}$ can induce $|\Delta_i| \gg 1$. However, such mass splitting can only decrease the unification scale, as shown below, and strengthen the constraints on the parameter space. We conclude that in the viable parameter space of the minimal setup, $\Delta$ is $\mathcal{O}(1)$. $\Delta = \mathcal{O}(10)$ can be achieved by adding more $SO(10)$ multiplets with split masses.

\begin{figure*}[t]
 \centering
        \begin{subfigure}[b]{0.49\textwidth}
         \centering
         \includegraphics[width=\textwidth]{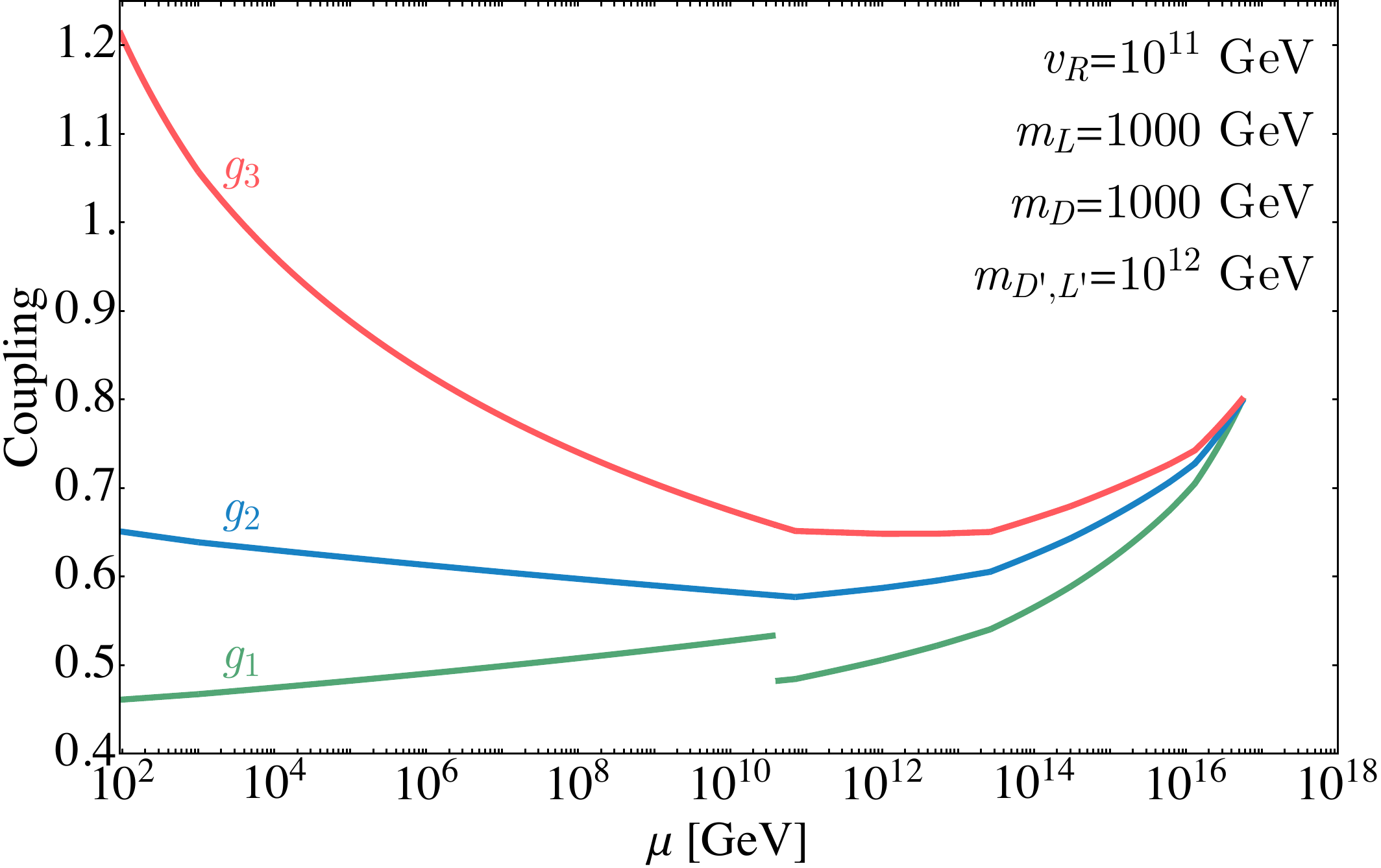}
         \caption{No mass splitting $m_D=m_L$}
     \end{subfigure}
     \begin{subfigure}[b]{0.49\textwidth}
         \centering
         \includegraphics[width=\textwidth]{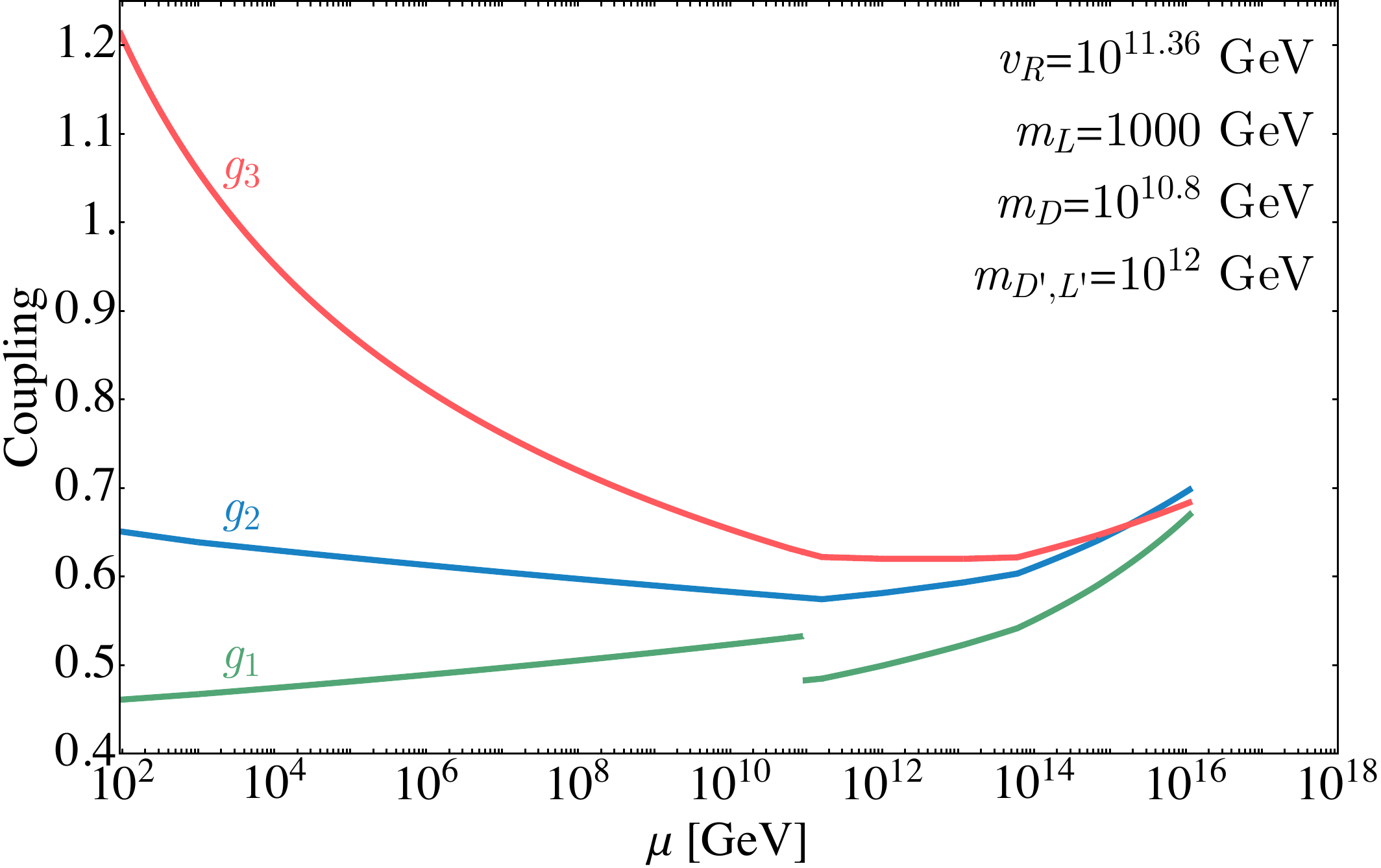}
         \caption{Mass splitting $m_D \gg m_L$}
         \end{subfigure}
    \caption{Precise gauge coupling unification for $1\textrm{ TeV}$ dark matter with $r_{XY}=2$ for (a) no mass splitting and (b) mass splitting, between $\chi_L$ and $\chi_D$.}
    \label{fig:RGEs}
\end{figure*}

Examples of the gauge coupling running are given in Fig.~\ref{fig:RGEs} for $m_{L}=1\textrm{ TeV}$, $r_{XY}=2$ and $m_{L',D'}= 10^{12}$ GeV. The choice of $m_{L'}= m_{D'}$ does not affect the unification at the 1-loop level. In the left panel, $m_D = 1$ TeV, for which $v_R=10^{11}$ GeV gives $\Delta(M_{XY}=10^{17}~\textrm{GeV})=0$. In the right panel, we take $m_{D}=10^{10.8}$ GeV, for which $v_{R}=10^{11.4}$ GeV gives the smallest $\Delta(M_{XY}=10^{16.1}~\textrm{GeV})=4$. The preferred unification scale is lower than the case with $m_{L}=m_D$.

\begin{figure*}[t]
 \centering
        \begin{subfigure}[b]{0.49\textwidth}
         \centering
         \includegraphics[width=\textwidth]{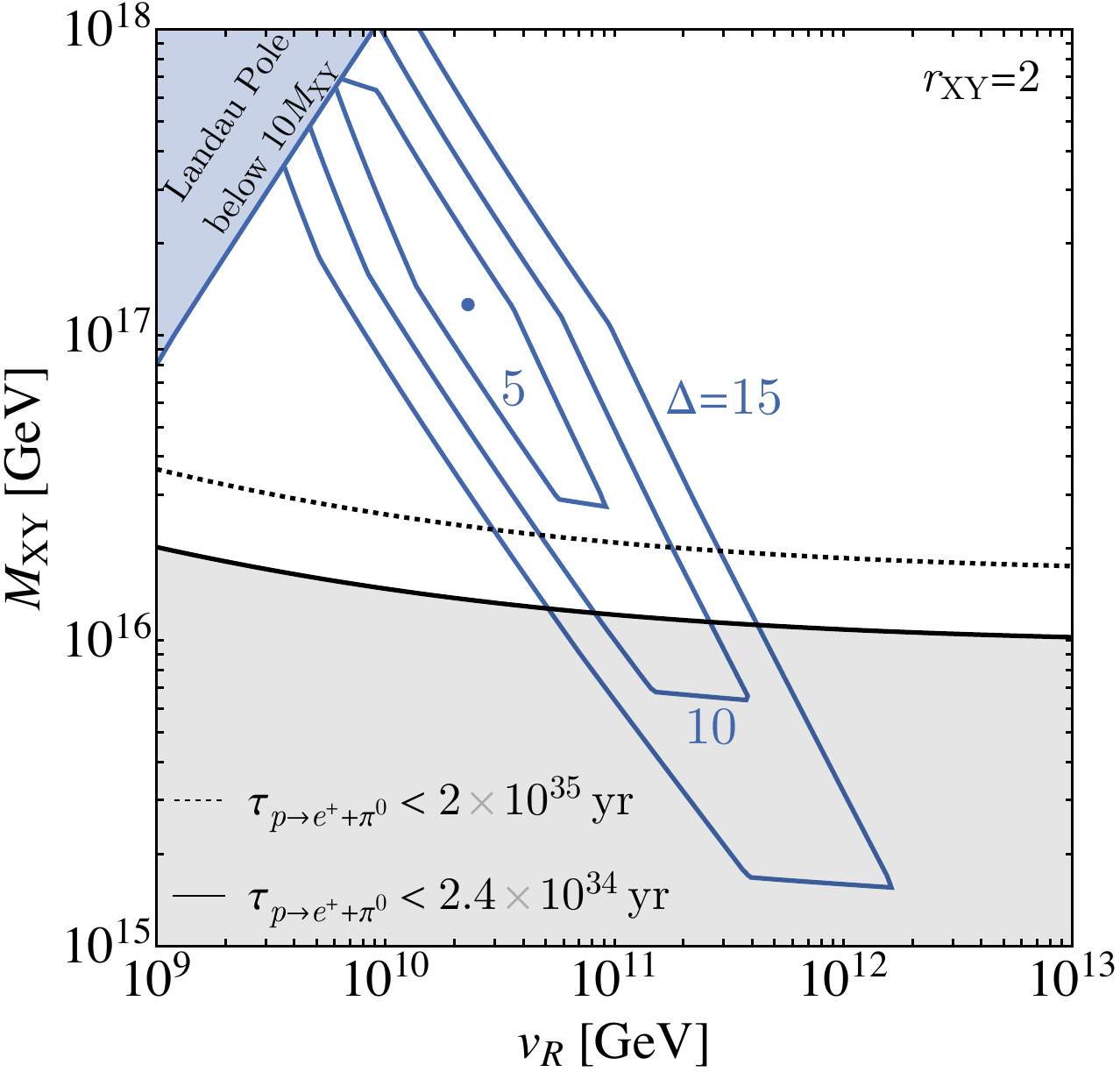}
         \caption{No dark matter, $r_{XY}=2$}
         \label{subfig:noDMr2}
     \end{subfigure}
     \begin{subfigure}[b]{0.49\textwidth}
         \centering
         \includegraphics[width=\textwidth]{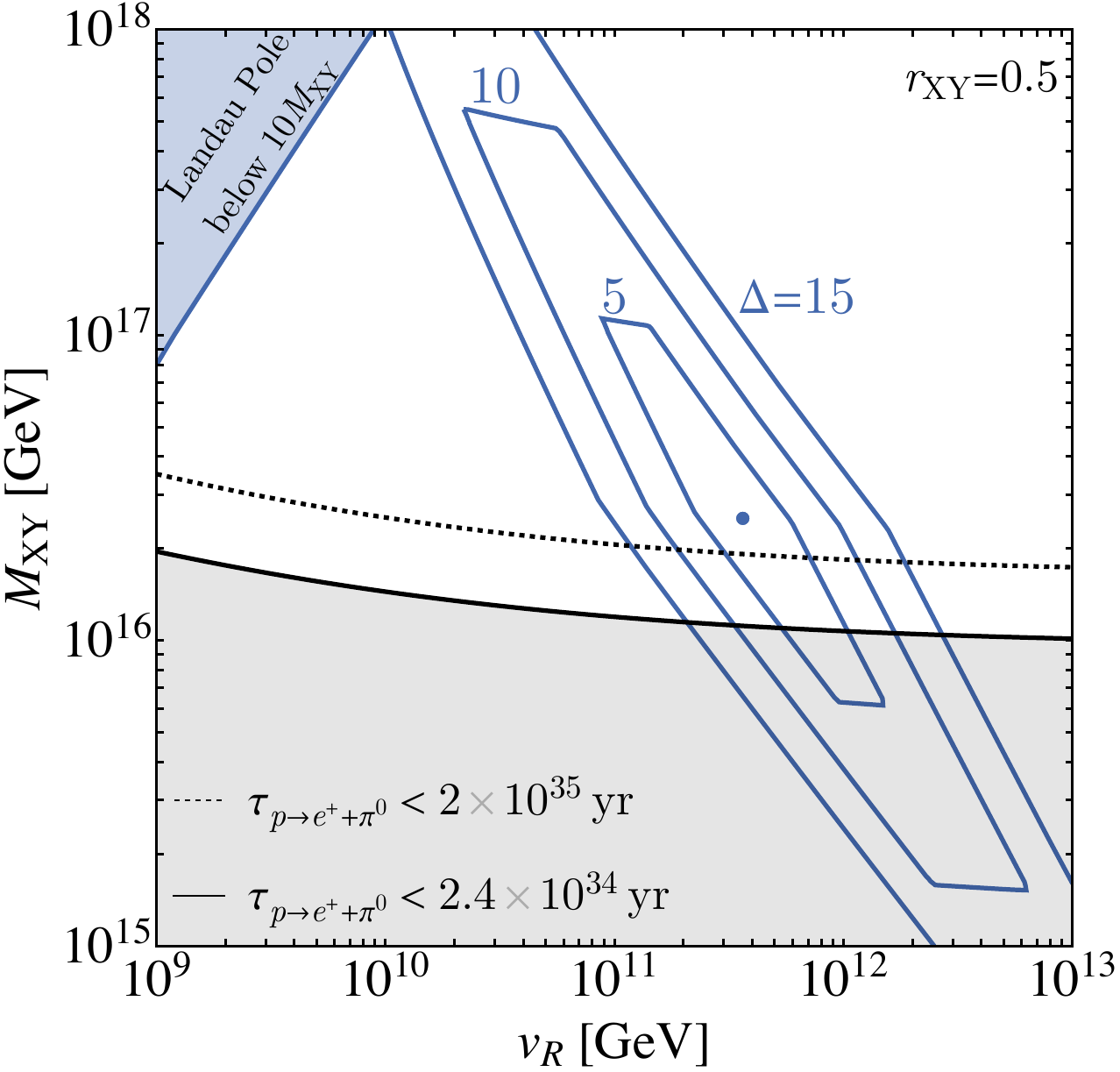}
         \caption{No dark matter, $r_{XY}=1/2$}
         \label{subfig:noDMr1o2}
         \end{subfigure}
                   \begin{subfigure}[b]{0.49\textwidth}
         \centering
         \includegraphics[width=\textwidth]{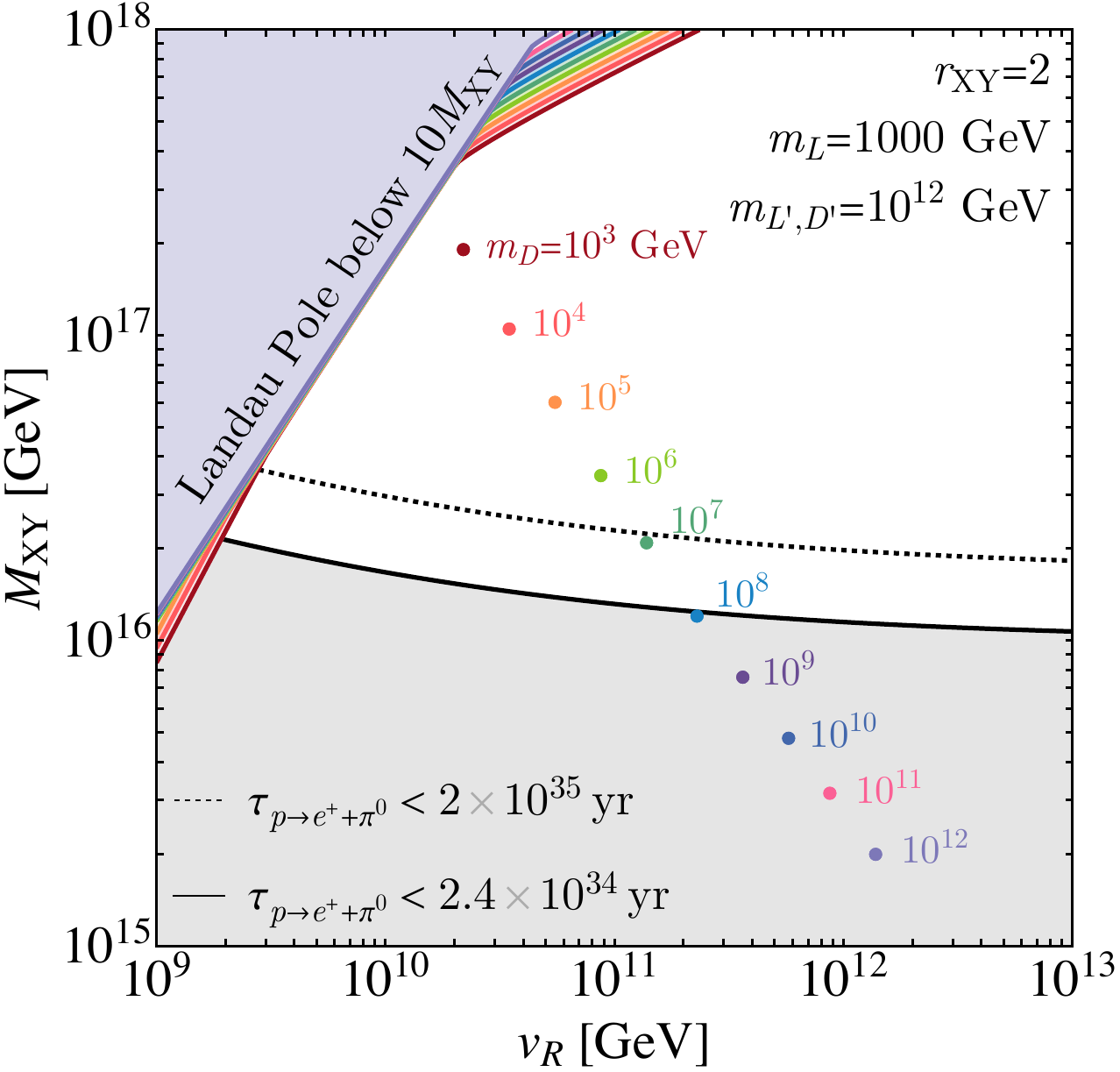}
         \caption{$m_L=1$ TeV, $r_{XY}=2$, $\Delta=0$}
         \label{subfig:mDrange1TeVr2}
     \end{subfigure}
            \begin{subfigure}[b]{0.49\textwidth}
         \centering
         \includegraphics[width=\textwidth]{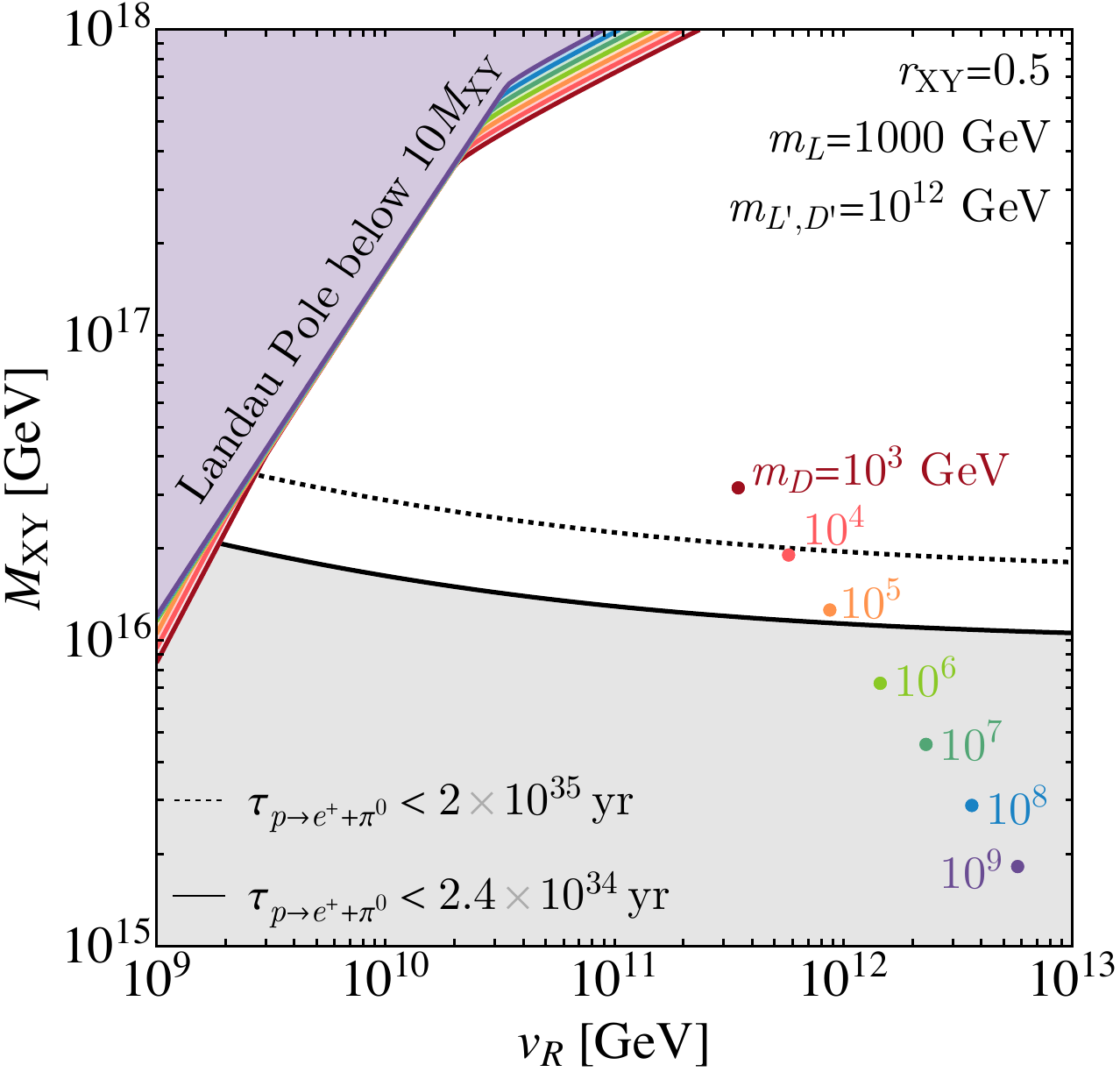}
         \caption{$m_L=1$ TeV, $r_{XY}=1/2$, $\Delta=0$}
         \label{subfig:mDrange1TeVr1o2}
     \end{subfigure}
    \caption{(a) and (b): required threshold corrections for precise gauge coupling unification in the $(v_R,M_{XY})$ plane without dark matter multiplets. (c) and (d): the effect of mass splitting between $\chi_L$ and $\chi_D$ on the $\Delta=0$ point. Larger mass splitting favours larger $v_R$ and smaller $M_{XY}$. The proton decay bounds in (c) and (d) correspond to $m_D=10^8$ GeV.}
    \label{fig:DeltaContours}
\end{figure*}

Fig.~\ref{fig:DeltaContours} shows $\Delta$ on the $(v_R,M_{XY})$ plane. Figs.~\ref{subfig:noDMr2} and~\ref{subfig:noDMr1o2} show the contours of $\Delta$ for the case without $\chi_{10_{1,2}}$ for $r_{XY}=2$ and $1/2$, respectively. Smaller $r_{XY}$ prefers larger $v_R$ and smaller $M_{XY}$. Figs.~\ref{subfig:mDrange1TeVr2} and~\ref{subfig:mDrange1TeVr1o2} show the points with $\Delta=0$ for the case with $\chi_{10_{1,2}}$ for $r_{XY}=2$ and $r_{XY}=1/2$, respectively. As $m_D$ increases, the preferred $v_R$ and $M_{XY}$ become larger and smaller, respectively. Proton decay bounds for Super-Kamiokande (SK) and the expected sensitivity of Hyper-Kamiokande (HK) are shown by gray-shaded regions and black-dotted lines, respectively. Smaller $M_{XY}$ leads to more rapid proton decay, as discussed in Sec.~\ref{sec:proton decay}. Together with the cosmological lower bound on $m_D$ in Eq.~\eqref{eq:mDlow}, the parameter space is strongly constrained, as discussed in Sec.~\ref{sec:cosmo_proton_constraints}. Possible mass splitting of $X_{45,j}$ by its coupling with $H_{45}$ also lowers the preferred $M_{XY}$. In the dark blue-shaded regions with low $v_R$ and large $\Delta$, labelled as ``Landau pole below $10M_{XY}$", the Landau pole scale of the gauge coupling constants is smaller than $10M_{XY}$. Higher-dimensional couplings between the $SO(10)$ gauge field and $H_{45}$, suppressed by the Landau pole scale, can give large tree-level threshold corrections to the gauge couplings and the requirement of precise gauge coupling unification, that is, small $\Delta$, does not make sense.

\subsection{Proton decay}\label{sec:proton decay}

Proton decay is induced generically in GUTs by $B$- and $L$-violating dimension-6 operators that are obtained by integrating out the heavy GUT-scale gauge bosons~\cite{Wilczek:1979hc,Weinberg:1979sa}. For the symmetry breaking chain we consider, the heavy $XY$ gauge bosons are integrated out to obtain dimension-6 operators in $G_{LR}$ and $G_{SM}$ that induce proton decay $p\rightarrow e^++\pi^0$. A similar analysis is performed in~\cite{Hamada:2020isl}. After integrating out the $XY$ gauge bosons, we obtain the $G_{SM}$ effective Lagrangian responsible for proton decay,
\begin{equation}
\begin{split}
    \mathcal{L}&=\frac{g_{10}^2}{M_{XY}^2}\left[2A_L (q\ell)(\bar{u}\bar{d})^\dag+A_R (qq)(\bar{u}\bar{e})^\dag\right] +\textrm{h.c.}\\
    &\supset\frac{g_{10}^2}{M_{XY}^2}\left[2A_L \left(ud\right)_Ru_Le_L+A_R \left(ud\right)_Lu_Re_R\right] +\textrm{h.c.},
\end{split}
\end{equation}
where the first line is written with left-handed Weyl fermions while the second line is written with Dirac fermions projected onto left- or right-handed components. 1-loop renormalization factors $A_{R,L}$ are obtained in terms of the fine-structure constants $\alpha_i$ and anomalous dimensions of the effective operators of $G_{LR}$ and $G_{SM}$ via RGE by taking $A_{R,L}(\Lambda_\text{GUT})=1$ at $\Lambda_\text{GUT}\approx M_{XY}=10^{15}\text{ GeV}$. $A_{R,L}$ can be written as
\begin{equation}
    A_{R,L}=A_{R,L}^{SM}\times A_{R,L}^{LR},
\end{equation}
where $A_{R,L}^{SM (LR)}$ is the $G_{SM (LR)}$ contribution given by~\cite{Caswell:1982fx}
\begin{equation}
    \begin{split}
        &A_{R}^{SM}=\prod_{n}\left(\frac{\alpha_3(\mu_{n+1})}{\alpha_3(\mu_{n})}\right)^{-\frac{2}{b^n_3}}\left(\frac{\alpha_2(\mu_{n+1})}{\alpha_2(\mu_{n})}\right)^{-\frac{9}{4b^n_2}}\left(\frac{\alpha_1(\mu_{n+1})}{\alpha_1(\mu_{n})}\right)^{-\frac{11}{12b^n_1}},\\
        &A_{L}^{SM}=\prod_{n}\left(\frac{\alpha_3(\mu_{n+1})}{\alpha_3(\mu_{n})}\right)^{-\frac{2}{b^n_3}}\left(\frac{\alpha_2(\mu_{n+1})}{\alpha_2(\mu_{n})}\right)^{-\frac{9}{4b^n_2}}\left(\frac{\alpha_1(\mu_{n+1})}{\alpha_1(\mu_{n})}\right)^{-\frac{23}{12b^n_1}},\\
        &A_{R}^{LR}=\prod_{n}\left(\frac{\alpha_3(\mu_{n+1})}{\alpha_3(\mu_{n})}\right)^{-\frac{2}{b^n_3}}\left(\frac{\alpha_2(\mu_{n+1})}{\alpha_2(\mu_{n})}\right)^{-\frac{9}{2b^n_2}}\left(\frac{\alpha_1(\mu_{n+1})}{\alpha_1(\mu_{n})}\right)^{-\frac{1}{4b^n_1}},\\
        &A_{L}^{LR}=A_{R}^{LR},
    \end{split}
\end{equation}
with index $n$ labeling the renormalization scale above which the 1-loop $\beta$-function coefficients are $b^n_{i}$, given in Appendix~\ref{app:BR_beta_fn}; see Sec.~\ref{sec:gauge running}.

The proton decay rate is given by
\begin{equation}
    \tau_{p\rightarrow e^+ +\pi^0}=\left[\frac{1}{32\pi}m_p\left(1-\frac{m_{\pi^0}^2}{m_p^2}\right)^2\frac{g_{10}^4}{M_{XY}^4}(4A_L^2+A_R^2)\abs{W_0}^2\right]^{-1},
\end{equation}
where $m_p$ ($m_{\pi^0}$) is the proton (pion) mass, and $W_0=-0.131\text{ GeV}^2$ is the pion-proton form factor at the renormalization scale $2\text{ GeV}$, obtained from lattice simulations, with a statistical uncertainty of 3.0\% and a systematic uncertainty of 9.7\%~\cite{Aoki:2017puj}.

The current experimental bound on the $p\rightarrow e^++\pi^0$ lifetime from SK is $\tau_{p\rightarrow e^++\pi^0}> 2.4\times 10^{34}$~years~(90\%~CL)~\cite{Super-Kamiokande:2020wjk}. HK will improve this bound to $\tau_{p\rightarrow e^++\pi^0}> 2\times 10^{35}$~years (90\%~CL)~\cite{Hyper-Kamiokande:2018ofw} if no proton decay is observed over 20 years of operation. The sensitivity of SK and HK on the unification scale are shown in Figs.~\ref{fig:DeltaContours},~\ref{fig:1TeVrXY2Contours},~\ref{fig:100GeVrXY2Contours},~\ref{fig:1TeVrXY1o2Contours},~\ref{fig:100GeVrXY1o2Contours} by gray-shaded regions and black-dotted lines, respectively. The lower bound on $M_{XY}$ becomes stronger for smaller $v_R$ because of lighter $X$-states that enhance the gauge coupling constants at high energy scales.

\subsection{Constraints on $v_R$ and $m_D$}\label{sec:cosmo_proton_constraints}

\begin{figure*}[t]
    \centering
     \begin{subfigure}[b]{0.49\textwidth}
         \centering
         \includegraphics[width=\textwidth]{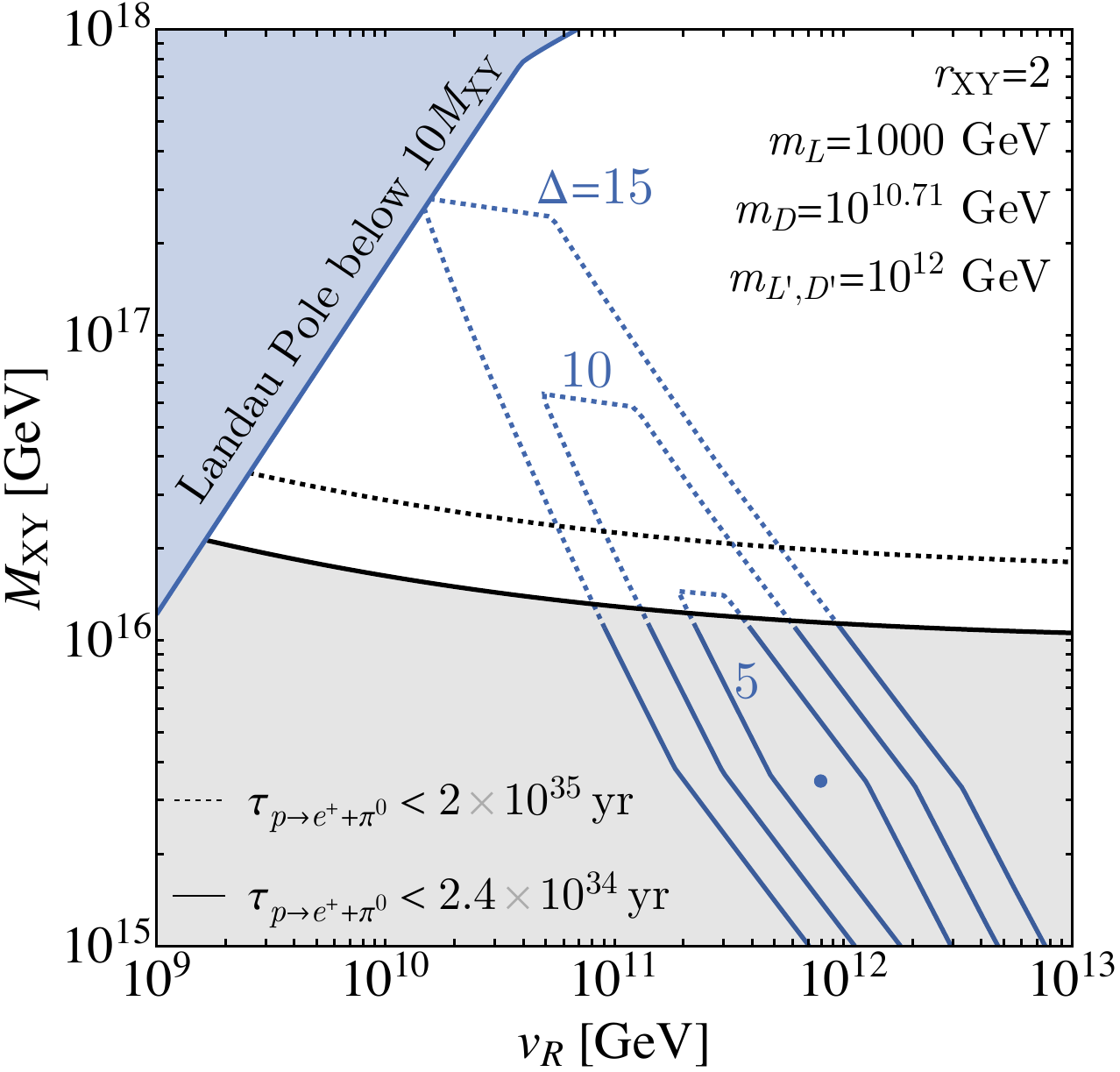}
        \caption{Lowest $m_D$ that satisfies SK constraint}
         \label{fig:1TeVSK}
     \end{subfigure}
          \begin{subfigure}[b]{0.49\textwidth}
         \centering
         \includegraphics[width=\textwidth]{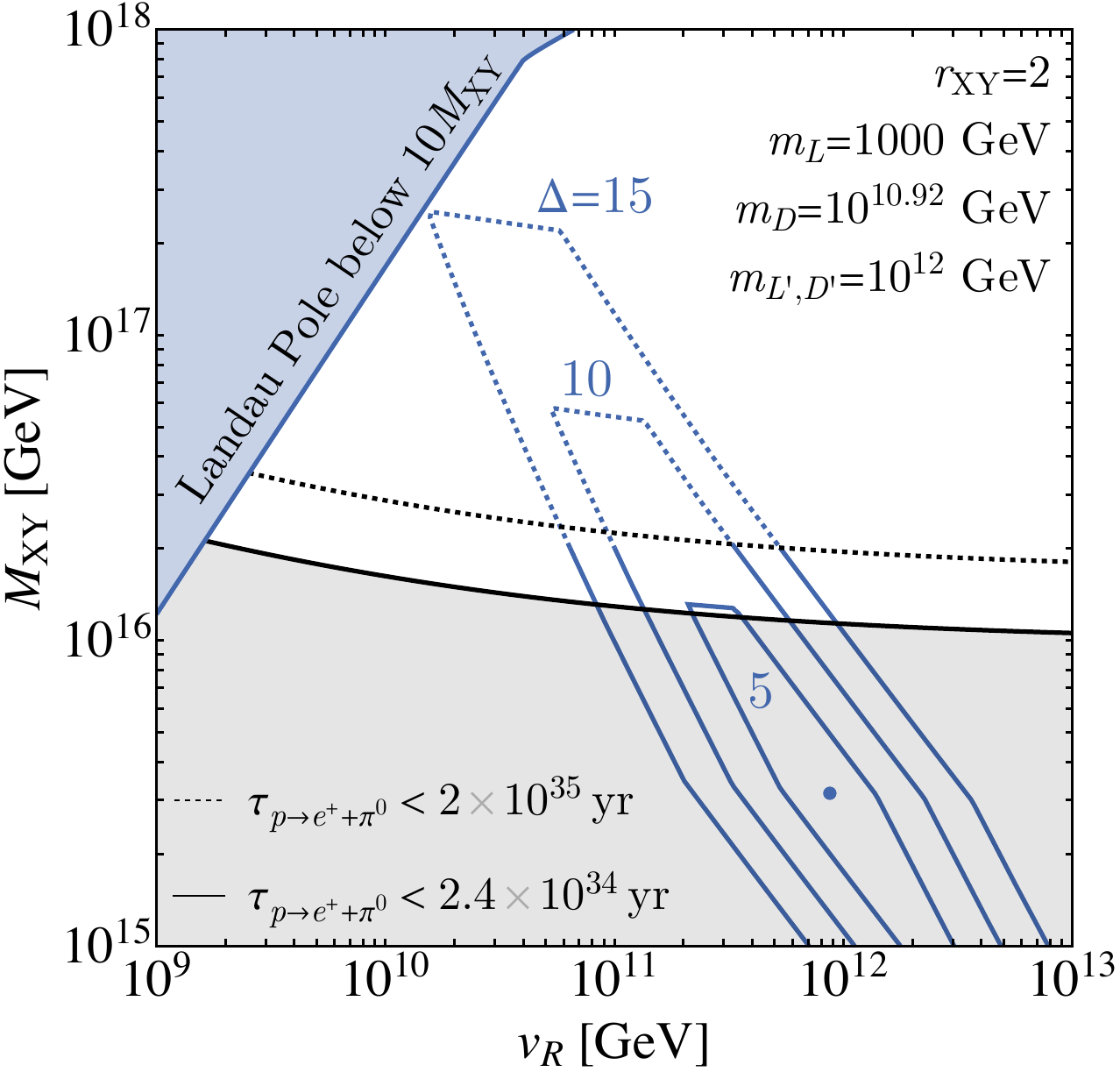}
     \caption{Lowest $m_D$ that satisfies HK constraint}
         \label{fig:1TeVHK}
         \end{subfigure}
     \begin{subfigure}[b]{0.49\textwidth}
    \centering
\includegraphics[width=\textwidth]{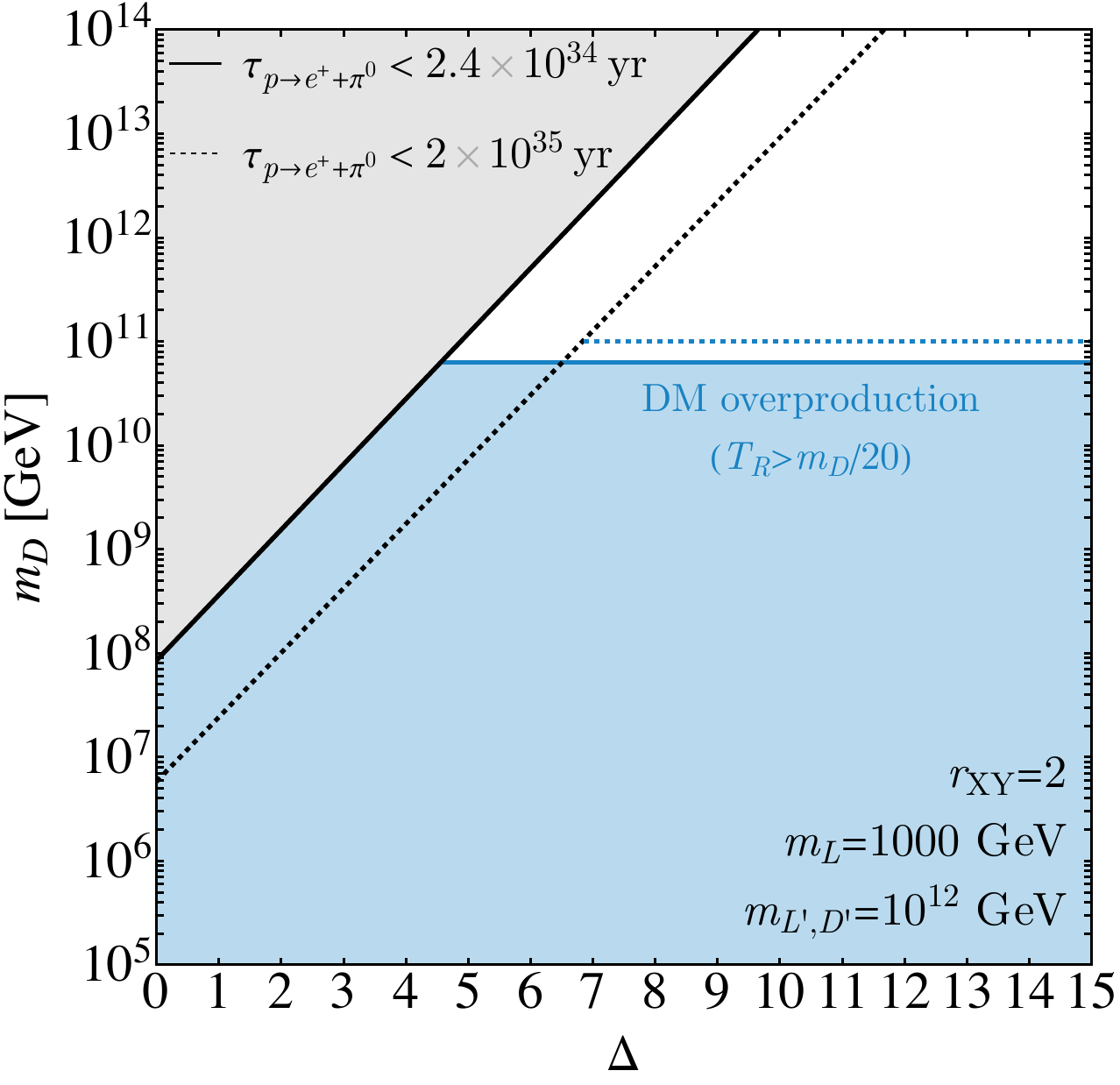}
    \caption{Constraint on $m_D$}
    \label{fig:mD_constraint_plot}
     \end{subfigure} 
         \begin{subfigure}[b]{0.49\textwidth}
    \centering
    \includegraphics[width=\textwidth]{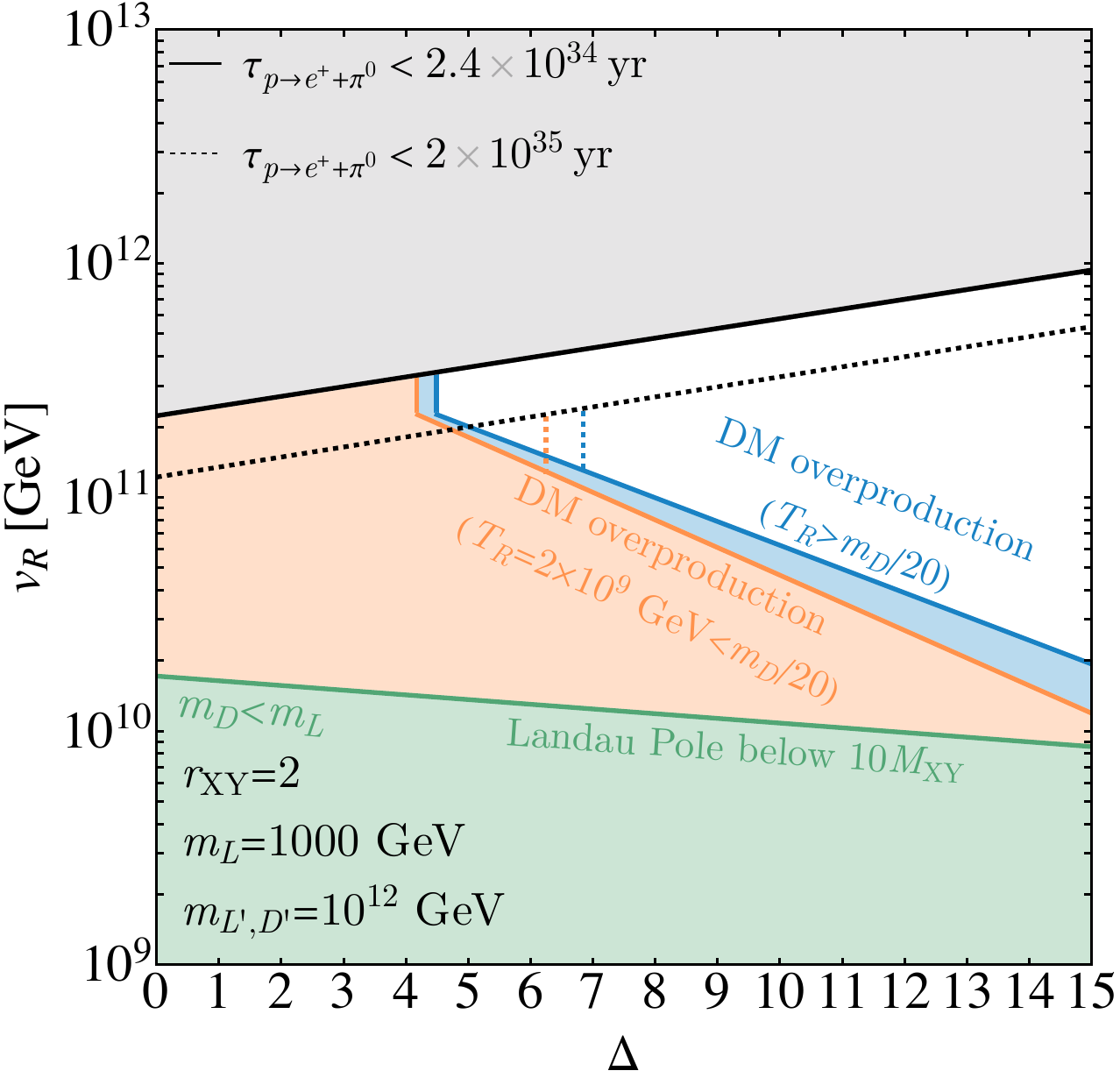}
    \caption{Constraint on $v_R$}
    \label{fig:vR_constraint_plot}
     \end{subfigure}
    \caption{
    Constraints on $m_D$ and $v_R$ for $m_L=1$ TeV and $r_{XY}=2$. (a) shows the contours of $\Delta$ for the smallest $m_D$ such that the $\Delta=15$ contour is consistent with the cosmological and SK proton decay bounds. (b) is an analogous plot with the prospect of HK. (c) and (d) show the viable range of $m_D$ and $v_R$, respectively, for a given $\Delta$.}
    \label{fig:1TeVrXY2Contours}
\end{figure*}

\begin{figure*}[t]
    \centering
     \begin{subfigure}[b]{0.49\textwidth}
         \centering
         \includegraphics[width=\textwidth]{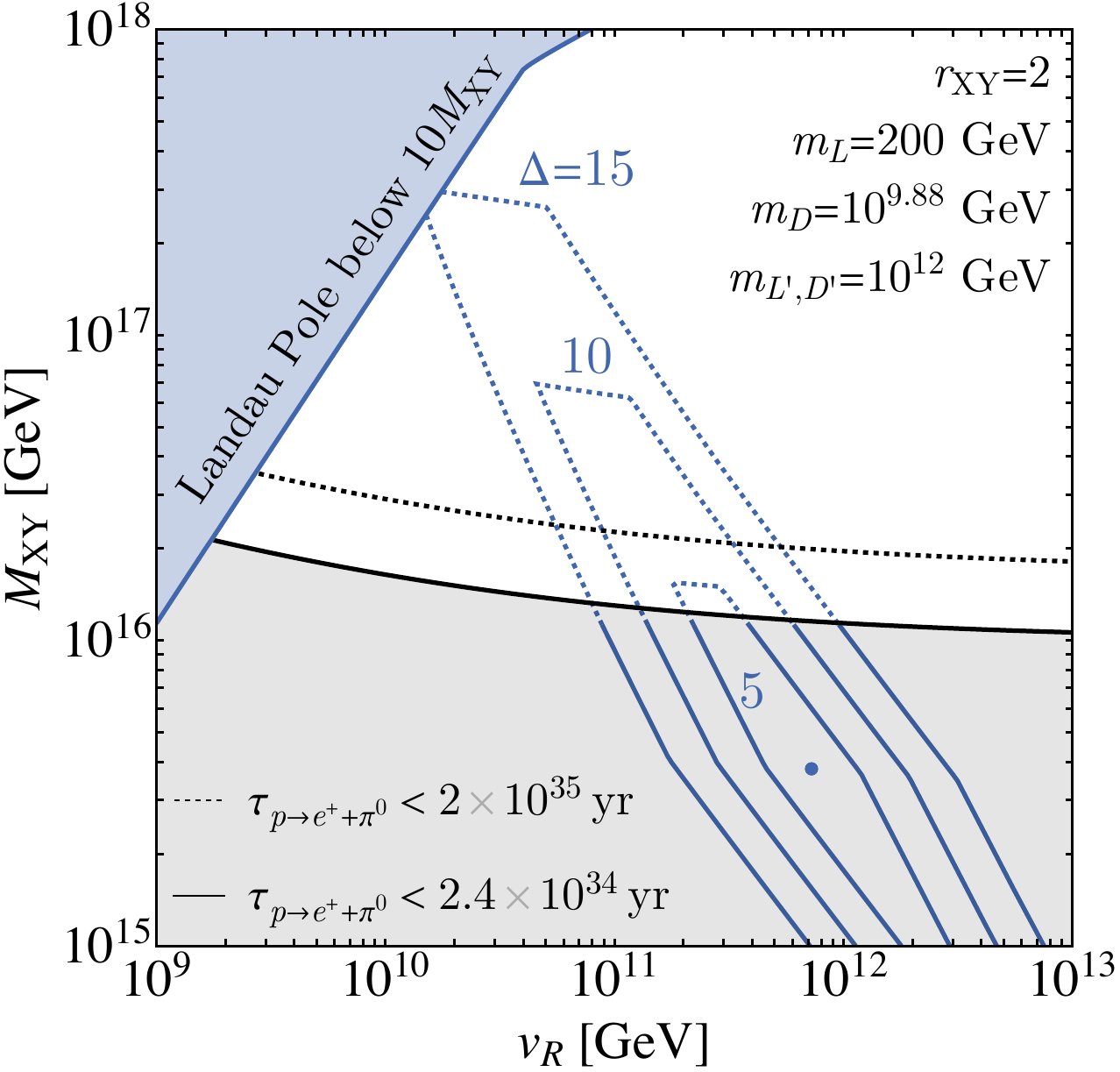}
         \caption{Lowest $m_D$ that satisfies SK constraint}
         \label{fig:100GeVSK}
     \end{subfigure}
     \begin{subfigure}[b]{0.49\textwidth}
         \centering
         \includegraphics[width=\textwidth]{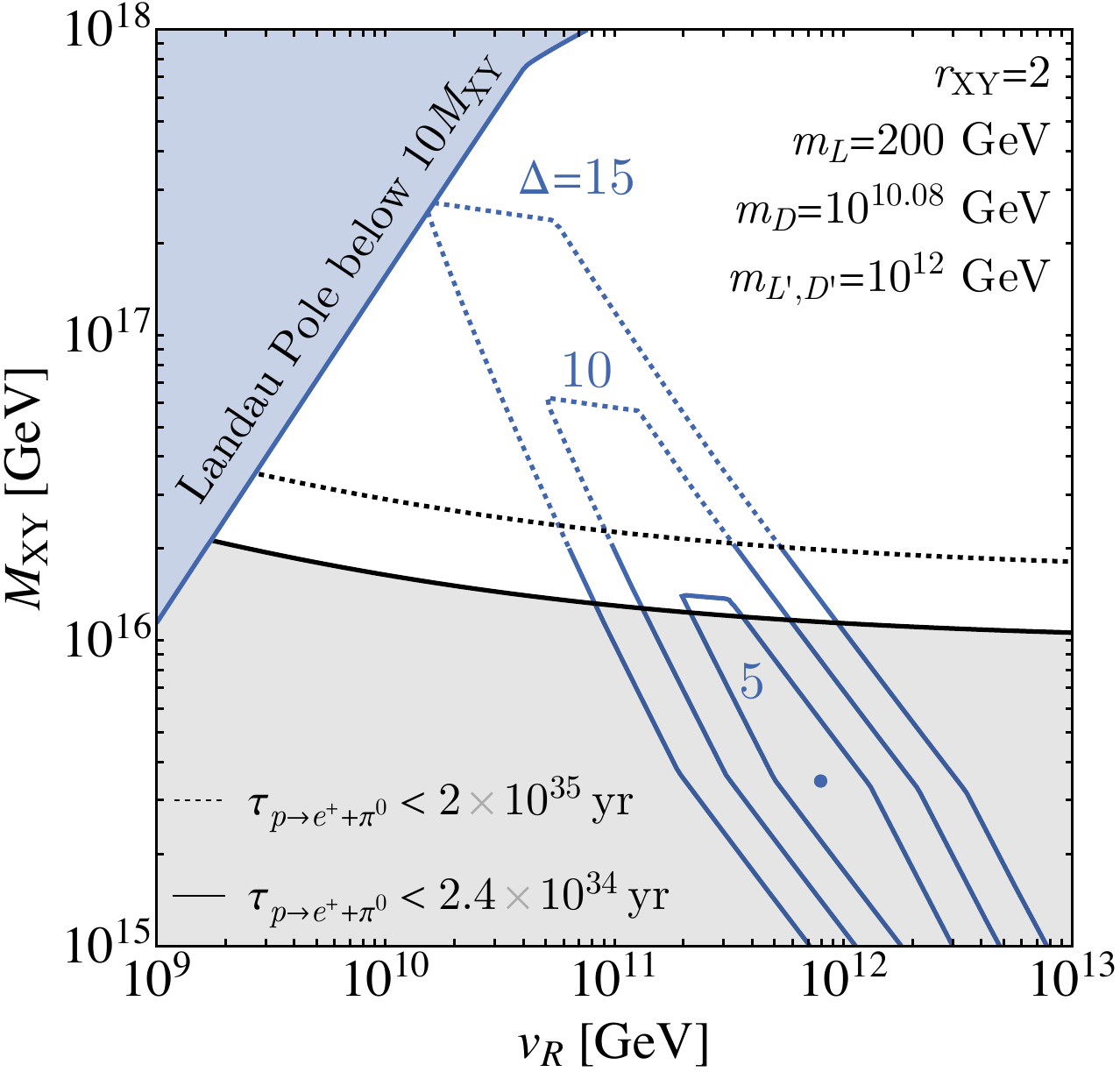}
         \caption{Lowest $m_D$ that satisfies HK constraint}
         \label{fig:100GeVHK}
         \end{subfigure}
        \begin{subfigure}[b]{0.49\textwidth}
    \centering
\includegraphics[width=\textwidth]{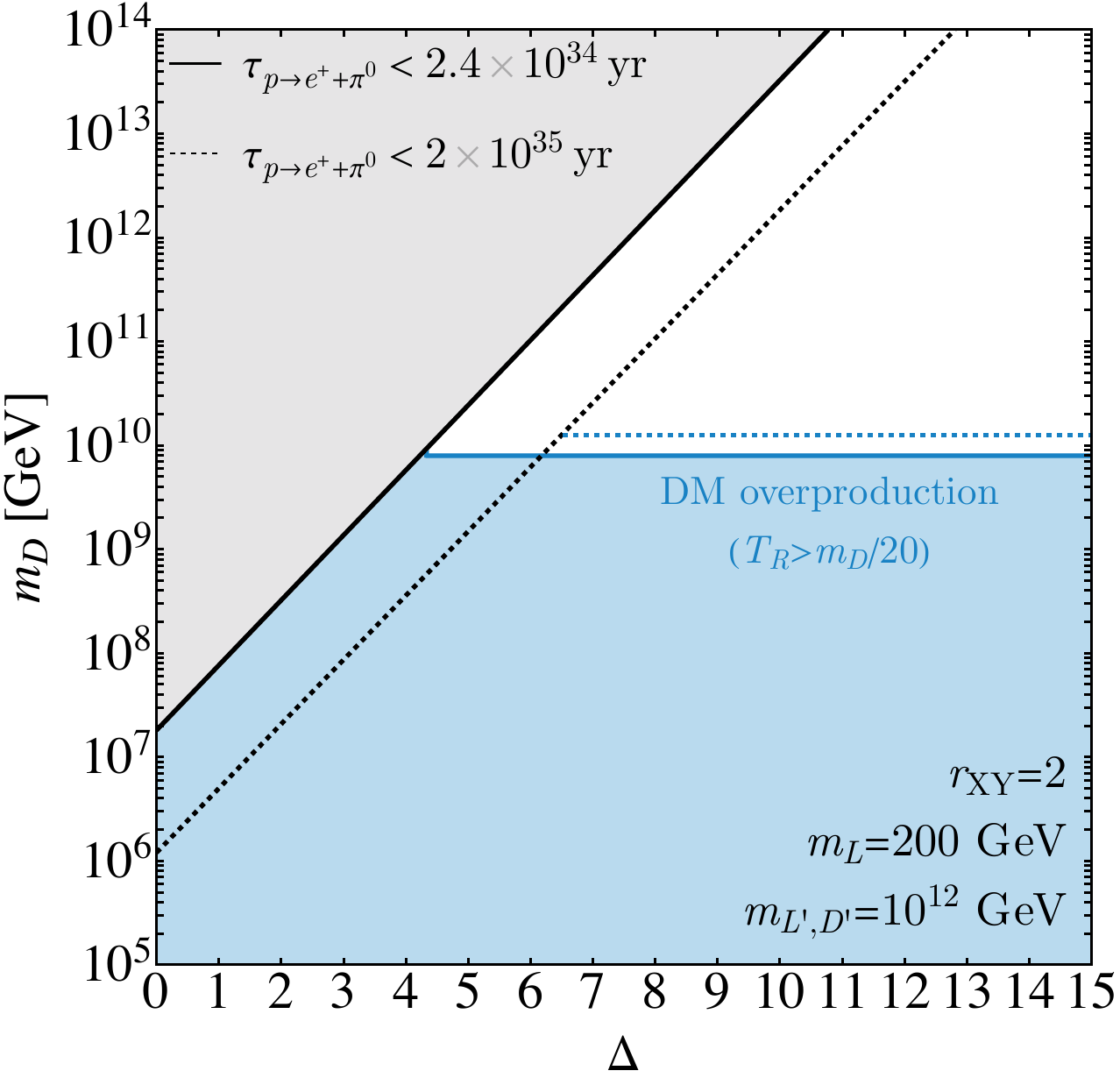}
    \caption{Constraint on $m_D$}
    \label{fig:mD_constraint_plot100}
     \end{subfigure}
     \begin{subfigure}[b]{0.49\textwidth}
    \centering
    \includegraphics[width=\textwidth]{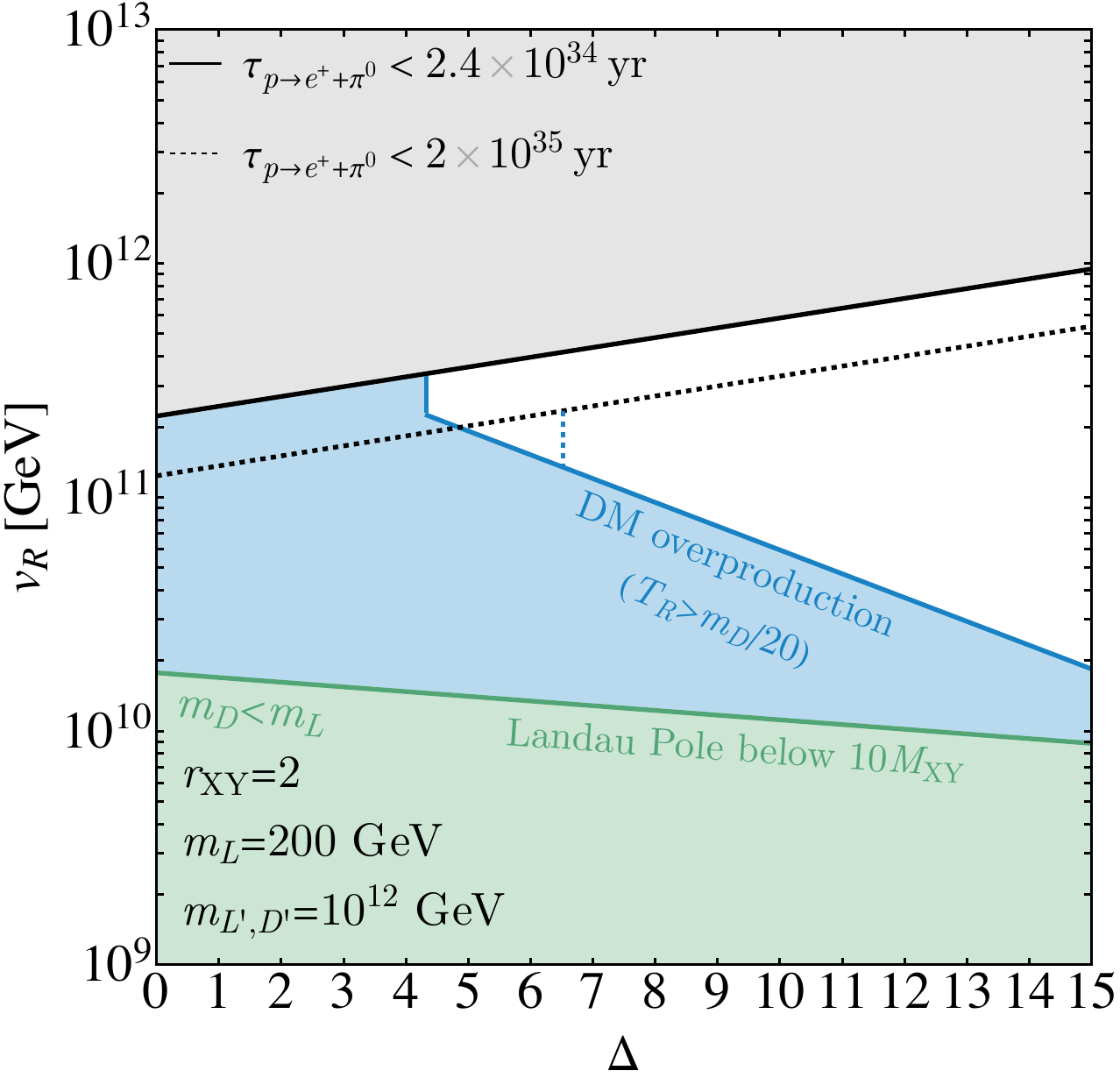}
    \caption{Constraint on $v_R$}
    \label{fig:vR_constraint_plot100}
     \end{subfigure}
    \caption{
    Same as Fig.~\ref{fig:1TeVrXY2Contours} with $m_L=200$ GeV.}
\label{fig:100GeVrXY2Contours}
\end{figure*}

We can now put together the cosmological bound in Sec.~\ref{sec:decay_of_colour_particles}, the gauge coupling unification in Sec.~\ref{sec:gauge running}, and the proton decay bound in Sec.~\ref{sec:proton decay} to restrict the viable range of $m_D$ and $v_R$. 

Fig.~\ref{fig:1TeVSK} shows the contours of $\Delta$ for $m_L=1$ TeV and $r_{XY}=2$ for the smallest $m_D$ such that the cosmological bound in Eq.~\eqref{eq:mDlow} is compatible with the SK proton decay bound when $\Delta=15$. Solid contour lines satisfy the cosmological bound while dotted contour lines do not. Fig.~\ref{fig:1TeVHK} shows the analogous plot for the expected sensitivity of HK. 

In Fig.~\ref{fig:mD_constraint_plot} we show the bound on $m_D$ for a given $\Delta$. As $m_D$ increases, the contours in Figs.~\ref{fig:1TeVSK} and~\ref{fig:1TeVHK} move toward the bottom-right (see Fig.~\ref{subfig:mDrange1TeVr2}), so in order to evade the proton decay bound, a larger $\Delta$ is required. The grey-shaded region and black-dotted diagonal line in Fig.~\ref{fig:mD_constraint_plot} correspond to this bound for SK and HK, respectively. Note that this bound comes solely from the proton decay bound and is applicable even if the cosmological bound is avoided by a low reheating temperature. For smaller $m_D$, the cosmological bound in Eq.~\eqref{eq:mDlow} requires smaller $M_{XY}$ and the proton decays too rapidly. The blue-shaded region and blue-dotted horizontal line correspond to this bound for SK and HK, respectively. The values of $m_D$ in Figs.~\ref{fig:1TeVSK} and~\ref{fig:1TeVHK} saturate this bound at $\Delta=15$.

In Fig.~\ref{fig:vR_constraint_plot} we show the bound on $v_R$ for a given $\Delta$. The upper-shaded region corresponds to the SK proton decay bound. The green lower-shaded region bounds $v_R$ from below as follows. For small $\Delta$, $v_R$ is bounded below by the requirement that $m_D>m_L$. For larger $\Delta$, in addition to requiring that $m_D>m_L$, the contours enter the region of $(v_R,M_{XY})$ where $M_{XY}$ is too close to the Landau pole scale for precise gauge unification to make sense. For these larger values of $\Delta$, the minimum $v_R$ lies on the boundary of the Landau pole constraint in the $(v_R,M_{XY})$ plots. The upper- and green lower-shaded regions are independent from the cosmological bound. The blue-shaded region corresponds to the combination of the SK proton decay bound with the cosmological bound in Eq.~\eqref{eq:mDlow}. If $T_R \ll m_D/20$, the cosmological bound in Eq.~\eqref{eq:mDlow} on $m_D$ and $M_{XY}$ can be avoided, and a wider range of $v_R$ is allowed. Still, if one requires successful thermal leptogenesis, the bound in Eq.~\eqref{eq:mDlowweak} is applicable. For $m_L=1$ TeV and $r_{XY}=2$, the bound happens to be similar to the lower bound on $m_D$ shown in Fig.~\ref{fig:mD_constraint_plot}. Still, since the bound on $M_{XY}$ in Eq.~\eqref{eq:mDlow} is lifted, the constraint on $v_R$ is relaxed, as shown by the orange-shaded region in Fig.~\ref{fig:vR_constraint_plot}. The dotted lines in Figs.~\ref{fig:mD_constraint_plot} and~\ref{fig:vR_constraint_plot} correspond to the same constraints but for the expected HK proton decay bound.

Fig.~\ref{fig:100GeVrXY2Contours} is the same as Fig.~\ref{fig:1TeVrXY2Contours}, but with $m_L=200$ GeV. The minimal required $\Delta$ and the prediction on $v_R$ are the same as those for $m_L=1$ TeV. For $m_L= 200$ GeV, however, the bound in Eq.~\eqref{eq:mDlowweak} is stronger than the lower bound on $m_D$ shown in Fig.~\ref{fig:mD_constraint_plot100}, and even though the bound on $M_{XY}$ is lifted, the bound on $v_R$ is not relaxed. Since the bound is not relaxed, we omit this constraint from Fig.~\ref{fig:vR_constraint_plot100}.

For $r_{XY}<2$, the contours of $\Delta$ on the $(v_R,M_{XY})$ plane
move toward the bottom-right (see Figs.~\ref{subfig:noDMr2} and~\ref{subfig:noDMr1o2}), so the preferred $v_R$ becomes larger while the proton decay constraint becomes stronger, and the required $\Delta$ becomes larger. See Appendix~\ref{sec:rXYhalf} for the figures with $r_{XY}=1/2$.

We comment on the possibility of dark matter in the ${\bf 45}$ or ${\bf 54}$ of $SO(10)$. The colored particles in those multiplets are subject to similar cosmological constraints as those on ${\bf 10}$ and a large mass splitting between dark matter and colored partners is required. Because the gauge coupling constant $\beta$-function contributions of ${\bf 45}$ and ${\bf 54}$ are larger than that of ${\bf 10}$, the mass splitting lowers the unification scale more than ${\bf 10}$ and the proton decays too rapidly.

\section{Standard Model Parameters}
\label{sec:SM}

As discussed in Sec.~\ref{sec:ParityBreaking}, the SM Higgs quartic coupling nearly vanishes at the Parity breaking scale up to calculable threshold corrections. We compute the running of the quartic coupling following~\cite{Buttazzo:2013uya}, adding the contribution of the dark matter multiplet to the running of the gauge coupling constants at the 1-loop level. The colored partner can also affect the running if $m_D < v_R$, but we find that the prediction on $v_R$ for the smallest allowed mass of the colored partner, $m_D=10^{10}$ GeV, differs from that for $m_D > v_R$ by less than 1\%.

\begin{figure}[t]
    \centering
    \includegraphics[width=0.6\textwidth]{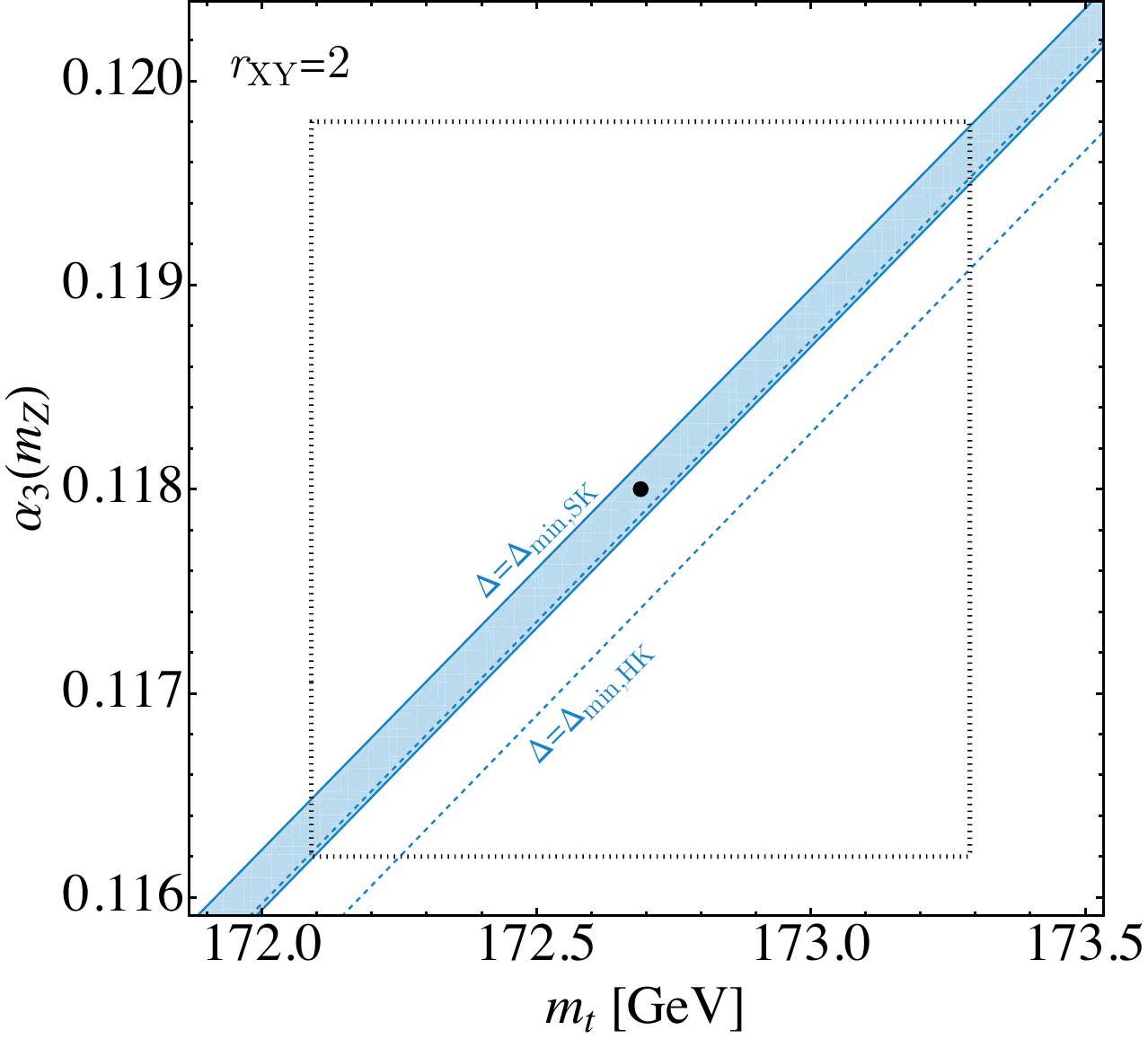}
    \caption{The prediction on the top quark mass $m_t$ and the strong coupling constant at the $Z$-boson mass $\alpha_3(m_Z)$. Here we take $r_{XY}=2$. In the blue-shaded region and between the blue-dashed lines, the required threshold correction is minimal for SK and HK, respectively. The black-dotted lines show the $2\sigma$ bound on $(m_t,\alpha_3(m_Z))$.}
    \label{fig:SMparameters}
\end{figure}

In Fig.~\ref{fig:SMparameters}, we show the prediction on the top quark mass $m_t$ and the strong coupling constant at the $Z$-boson mass $\alpha_3(m_Z)$ from precise gauge coupling unification, and the constraints from cosmological and proton decay bounds. The blue-shaded region and blue-dashed lines give the range of $v_R$ for the minimal value of $\Delta$, shown in Fig.~\ref{fig:vR_constraint_plot}, for SK and HK, respectively. One can see that the cosmological and proton decay bounds, together with precise gauge coupling unification, predicts $(m_t,\alpha_3(m_Z))$ in a narrow region. The dot and the rectangle with a dotted edge show the central value and $2\sigma$ allowed range of $(m_t,\alpha_3(m_Z))$, respectively~\cite{ParticleDataGroup:2022pth}, which is consistent with our prediction. Improved lattice computation and measurements of the $Z$-pole at future lepton colliders can determine $\alpha_3(m_Z)$ with an accuracy of $0.0001$~\cite{Lepage:2014fla,TLEPDesignStudyWorkingGroup:2013myl}. Future lepton colliders can also determine the top quark mass with an accuracy of a few 10 MeV~\cite{Seidel:2013sqa,Horiguchi:2013wra,Kiyo:2015ooa,Beneke:2015kwa} and test our prediction.

The threshold correction to the quartic coupling at $v_R$ from the top quark Yukawa is computed using the formulae derived in~\cite{Hall:2019qwx}, fixing the up-type Yukawa couplings to obtain the correct bottom/tau Yukawa ratio from the $SO(10)$ breaking in the masses of $X_{45,i}$ that affects the mixing between $\psi_i$ and $X_{45,i}$. See~\cite{Hall:2019qwx} for details. If the bottom/tau ratio is explained in a different way, the prediction on $m_t$ can become smaller, so the prediction on $m_t$ in Fig.~\ref{fig:SMparameters} can be understood as an upper bound on $m_t$.

\section{Summary}

The strong $CP$ problem can be solved by Parity symmetry with a left-right extended gauge group. The extended gauge group can be embedded into the $SO(10)$ unified gauge group. In this paper, we investigated an electroweak-charged dark matter candidate in the unified theory and its implications on precise gauge coupling unification.

Dark matter is taken to be a fermion with an $SU(2)_L\times U(1)$ charge of $({\bf 2},1/2)$. It has a colored $SO(10)$ partner which decays into dark matter via the exchange of heavy gauge bosons. In order for the colored partner to decay without overproducing dark matter, it should be much heavier than dark matter. Such a mass splitting can be naturally achieved by the coupling of an $SO(10)\times CP$ breaking Higgs in ${\bf 45}$ to the dark matter multiplet. We find that large mass splitting, via quantum corrections to the gauge coupling constants, lowers the preferred unification scale and enhances the proton decay rate. Super-Kamiokande has already excluded the parameter region with $\Delta <4$ and Hyper-Kamiokande will probe the parameter region with $\Delta <7$.

If the freeze-out mechanism determines the dark matter abundance, the dark matter mass should be 1 TeV. However, the decay of the colored partner can produce extra dark matter, and the dark matter mass may be as low as the LHC bound of 200 GeV, or, if the mass splitting between the charged and neutral components is sufficiently large, the LEP bound of 100 GeV. High-Luminosity LHC can probe the dark matter mass up to 500 GeV, and gamma-ray observations can detect the dark matter annihilation if the galactic center has a cuspy dark matter halo profile.

The model also has implications to the measurements of SM parameters. In the minimal Higgs model, the SM Higgs quartic coupling is predicted to vanish around the Parity symmetry breaking scale up to calculable threshold corrections. The Parity symmetry breaking scale is also determined by the requirement of precise gauge coupling unification. Since the running of the SM Higgs quartic coupling is sensitive to the top quark mass and the strong coupling constant, precise gauge coupling unification predicts the range of these two parameters. In the parameter region that requires minimal threshold correction to the gauge coupling constants, the top quark mass is predicted within the range of 100 MeV for a given strong coupling constant. The prediction can be confirmed by future lepton colliders.

\section*{Acknowledgement}

The work of K.H.~was supported by Grant-in-Aid for Scientific Research from the Ministry of Education, Culture, Sports, Science, and Technology (MEXT), Japan (20H01895), and by World Premier International Research Center Initiative (WPI), MEXT, Japan (Kavli IPMU).

\appendix
\section{Branching Rules and \boldmath$\beta$-function Coefficients}\label{app:BR_beta_fn}

In this appendix, we provide the relevant branching rules for $SO(10)\rightarrow G_{LR}\rightarrow G_{SM}$, and the 1-loop and 2-loop $\beta$-function coefficients $b_i$ and $b_{ij}$. The branching rules and $\beta$-function coefficients for the $\mathbf{10}$, $\mathbf{45}$, and $\mathbf{54}$ of $SO(10)\rightarrow G_{LR}\rightarrow G_{SM}$ are shown in Tables~\ref{tab:branching_rules_10}, ~\ref{tab:branching_rules_45}, and~\ref{tab:branching_rules_54}, respectively. In showing the $\beta$-function coefficients, we separate the contributions of colored particles from those of non-colored particles.

\begin{table}
    \centering
    \begin{tabular}{|c||c|c|c|c|}
 \hline
 $SO(10)$&\multicolumn{4}{|c|}{$\mathbf{10}$}\\
 \hline
 & $D$ &$\bar{D}$ &\multicolumn{2}{|c|}{$\Delta$}\\
 \hline
 $SU(3)$&3&$\bar{3}$&\multicolumn{2}{|c|}{1}\\
 $SU(2)_L$&1&1&\multicolumn{2}{|c|}{2}\\
 $SU(2)_R$&1&1&\multicolumn{2}{|c|}{2}\\
 $U(1)$&-1/3&1/3&\multicolumn{2}{|c|}{0}\\
 \hline
  $b_i^\text{LR}$&\multicolumn{2}{|c|}{$\left(\begin{matrix}
        -2/3\\0\\-2/3
    \end{matrix}\right)$}&\multicolumn{2}{|c|}{$\left(\begin{matrix}
        0\\-2/3\\0
    \end{matrix}\right)$}\\
 \hline
 $b_{ij}^\text{LR}$&\multicolumn{2}{|c|}{$\left(\begin{matrix}
        -\frac{1}{6}&0&-\frac{4}{3}\\0&0&0\\-\frac{1}{6}&0&-\frac{19}{3}
    \end{matrix}\right)$}&\multicolumn{2}{|c|}{$\left(\begin{matrix}
        0&0&0\\0&-\frac{29}{6}&0\\0&0&0
    \end{matrix}\right)$}\\
 \hline
 \hline
 &$D$&$\bar{D}$&$\bar{L}$&$L$\\
 \hline
 $SU(3)$&3&$\bar{3}$& 1&1\\
 $SU(2)_L$&1&1& 2&2\\
 $U(1)$&-1/3&1/3& 1/2&-1/2\\
 \hline
 $b_i^\text{SM}$&\multicolumn{2}{|c|}{$\left(\begin{matrix}
        -4/15\\0\\-2/3
    \end{matrix}\right)$}&\multicolumn{2}{|c|}{$\left(\begin{matrix}
        -2/5\\-2/3\\0
    \end{matrix}\right)$}\\
 \hline
 $b_{ij}^\text{SM}$&\multicolumn{2}{|c|}{$\left(\begin{matrix}
        -\frac{2}{75}&0&-\frac{8}{15}\\0&0&0\\-\frac{1}{15}&0&-\frac{19}{3}
    \end{matrix}\right)$}&\multicolumn{2}{|c|}{$\left(\begin{matrix}
        -\frac{9}{100}&-\frac{9}{20}&0\\-\frac{3}{20}&-\frac{49}{12}&0\\0&0&0
    \end{matrix}\right)$}\\
 \hline
 
\end{tabular}
\caption{Branching rules and $\beta$-function coefficients of the $\mathbf{10}$ of $SO(10)$.}
\label{tab:branching_rules_10}
\end{table}

\begin{table}
    \centering

\begin{tabular}{|c||c|c|c|c|c|c|c|c|c|c|c|c|}
 \hline
 $SO(10)$&\multicolumn{12}{|c|}{$\mathbf{45}$}\\

 \hline
 $SU(3)$&8&3&$\bar{3}$&\multicolumn{2}{|c|}{3}&\multicolumn{2}{|c|}{$\bar{3}$}&1&\multicolumn{3}{|c|}{1}&1\\
 $SU(2)_L$&1&1&1&\multicolumn{2}{|c|}{2}&\multicolumn{2}{|c|}{2}&3&\multicolumn{3}{|c|}{1}&1\\
 $SU(2)_R$&1&1&1&\multicolumn{2}{|c|}{2}&\multicolumn{2}{|c|}{2}&1&\multicolumn{3}{|c|}{3}&1\\
 $U(1)$&0&2/3&-2/3&\multicolumn{2}{|c|}{-1/3}&\multicolumn{2}{|c|}{1/3}&0&\multicolumn{3}{|c|}{0}&0\\
 \hline
 $b_i^\text{LR}$&\multicolumn{7}{|c|}{$\left(\begin{matrix}
        -16/3\\-12/3\\-16/3
    \end{matrix}\right)$}&\multicolumn{5}{|c|}{$\left(\begin{matrix}
        0\\-4/3\\0
    \end{matrix}\right)$}\\
 \hline
 $b_{ij}^\text{LR}$&\multicolumn{7}{|c|}{$\left(\begin{matrix}
        -\frac{10}{3}&-6&-\frac{32}{3}\\-1&-29&-8\\-\frac{4}{3}&-6&-\frac{167}{3}
    \end{matrix}\right)$}&\multicolumn{5}{|c|}{$\left(\begin{matrix}
        0&0&0\\0&-\frac{32}{3}&0\\0&0&0
    \end{matrix}\right)$}\\
 \hline
 \hline
 $SU(3)$&8&3&$\bar{3}$&3&3&$\bar{3}$&$\bar{3}$&1&1&1&1&1\\
 $SU(2)_L$&1&1&1&2&2&2&2&3&1&1&1&1\\
 $U(1)$&0&2/3&-2/3&1/6&-5/6&-1/6&5/6&0&1&-1&0&0\\
 \hline
 $b_i^\text{SM}$&\multicolumn{7}{|c|}{$\left(\begin{matrix}
        -68/15\\-12/3\\-16/3
    \end{matrix}\right)$}&\multicolumn{5}{|c|}{$\left(\begin{matrix}
        -4/5\\-4/3\\0
    \end{matrix}\right)$}\\
 \hline
 $b_{ij}^\text{SM}$&\multicolumn{7}{|c|}{$\left(\begin{matrix}
        -\frac{377}{150}&-\frac{39}{10}&-\frac{136}{15}\\-\frac{13}{10}&-\frac{49}{2}&-8\\-\frac{17}{15}&-3&-\frac{167}{3}
    \end{matrix}\right)$}&\multicolumn{5}{|c|}{$\left(\begin{matrix}
        -\frac{18}{25}&0&0\\0&-\frac{32}{3}&0\\0&0&0
    \end{matrix}\right)$}\\
 \hline
\end{tabular}
\caption{Branching rules and $\beta$-function coefficients of the $\mathbf{45}$ of $SO(10)$.}
\label{tab:branching_rules_45}
\end{table}

\begin{table}
    \centering

\begin{tabular}{|c||c|c|c|c|c|c|c|c|c|c|c|}
 \hline
 $SO(10)$&\multicolumn{11}{|c|}{$\mathbf{54}$}\\
 \hline
 $SU(3)$&8&6&$\bar{6}$&\multicolumn{2}{|c|}{3}&\multicolumn{2}{|c|}{$\bar{3}$}&\multicolumn{3}{|c|}{1}&1\\
 $SU(2)_L$&1&1&1&\multicolumn{2}{|c|}{2}&\multicolumn{2}{|c|}{2}&\multicolumn{3}{|c|}{3}&1\\
 $SU(2)_R$&1&1&1&\multicolumn{2}{|c|}{2}&\multicolumn{2}{|c|}{2}&\multicolumn{3}{|c|}{3}&1\\
 $U(1)$&0&2/3&-2/3&\multicolumn{2}{|c|}{-1/3}&\multicolumn{2}{|c|}{1/3}&\multicolumn{3}{|c|}{0}&0\\
 \hline
  $b_i^\text{LR}$&\multicolumn{7}{|c|}{$\left(\begin{matrix}
        -8\\-4\\-8
    \end{matrix}\right)$}&\multicolumn{4}{|c|}{$\left(\begin{matrix}
        0\\-4\\0
    \end{matrix}\right)$}\\
 \hline
 $b_{ij}^\text{LR}$&\multicolumn{7}{|c|}{$\left(\begin{matrix}
        -6&-6&-32\\-1&-29&-8\\-4&-6&-91
    \end{matrix}\right)$}&\multicolumn{4}{|c|}{$\left(\begin{matrix}
        0&0&0\\0&-44&0\\0&0&0
    \end{matrix}\right)$}\\
 \hline
 \hline
 $SU(3)$&8&6&$\bar{6}$&3&$3$&$\bar{3}$&$\bar{3}$&1&1&1&1\\
 $SU(2)_L$&1&1&1&2&2&2&2&3&3&3&1\\
 $U(1)$&0&2/3&-2/3&1/6&-5/6&-1/6&5/6&1&-1&0&0\\
\hline
 $b_i^\text{SM}$&\multicolumn{7}{|c|}{$\left(\begin{matrix}
        -28/5\\-4\\-8
    \end{matrix}\right)$}&\multicolumn{4}{|c|}{$\left(\begin{matrix}
        -12/5\\-4\\0
    \end{matrix}\right)$}\\
 \hline
 $b_{ij}^\text{SM}$&\multicolumn{7}{|c|}{$\left(\begin{matrix}
        -\frac{147}{50}&-\frac{39}{10}&-\frac{88}{5}\\-\frac{13}{10}&-\frac{49}{2}&-8\\-\frac{11}{5}&-3&-91
    \end{matrix}\right)$}&\multicolumn{4}{|c|}{$\left(\begin{matrix}
        -\frac{54}{25}&-\frac{36}{5}&0\\-\frac{12}{5}&-32&0\\0&0&0
    \end{matrix}\right)$}\\
 \hline
\end{tabular}
    \caption{Branching rules and $\beta$-function coefficients of the $\mathbf{54}$ of $SO(10)$.}
\label{tab:branching_rules_54}
\end{table}

\clearpage

\section{Constraints for \boldmath$r_{XY}=1/2$}
\label{sec:rXYhalf}

\begin{figure*}[h]
    \centering

     \begin{subfigure}[b]{0.49\textwidth}
         \centering
         \includegraphics[width=\textwidth]{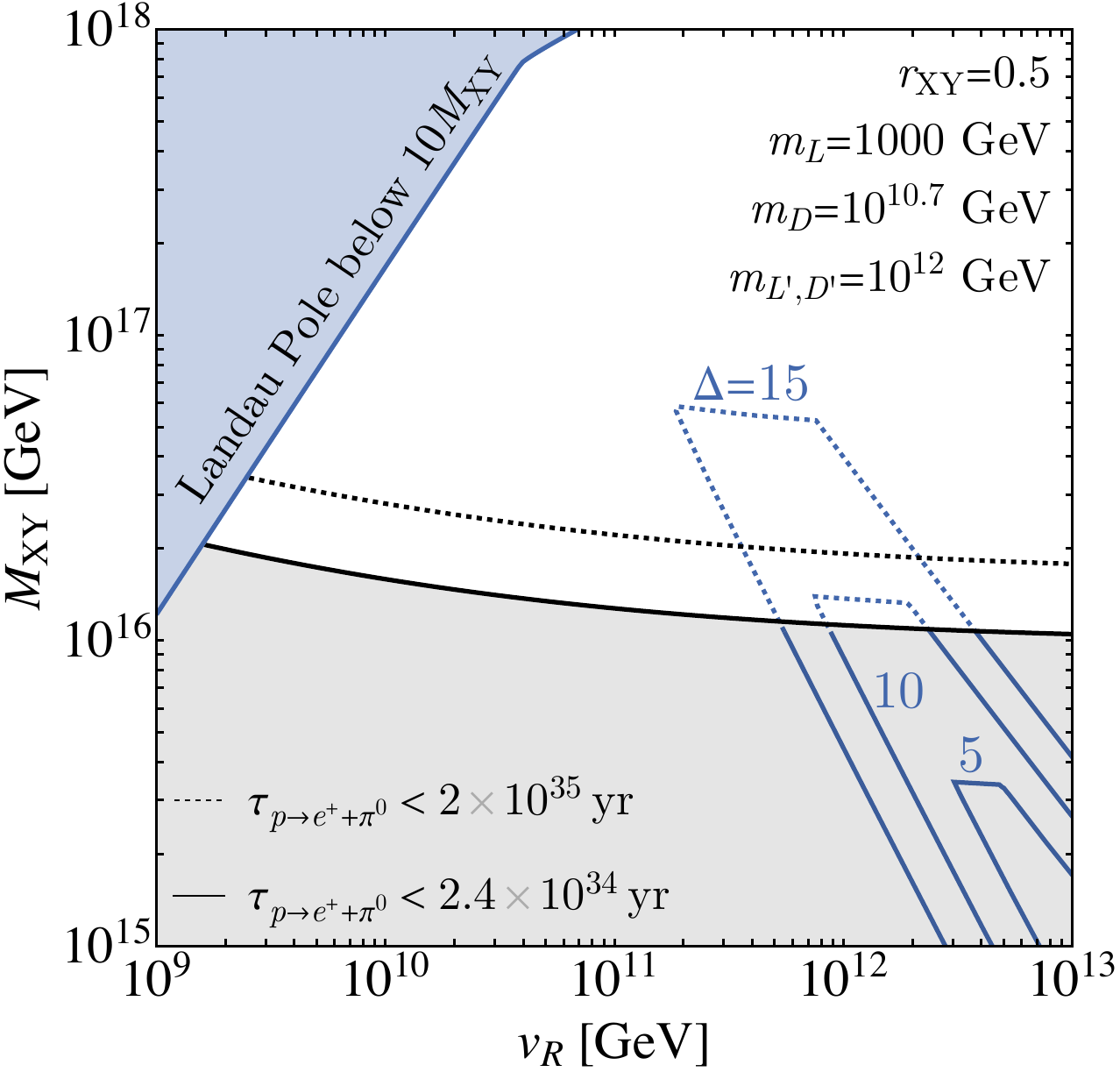}
         \caption{Lowest $m_D$ that satisfies SK constraint}
     \end{subfigure}
     \begin{subfigure}[b]{0.49\textwidth}
         \centering
         \includegraphics[width=\textwidth]{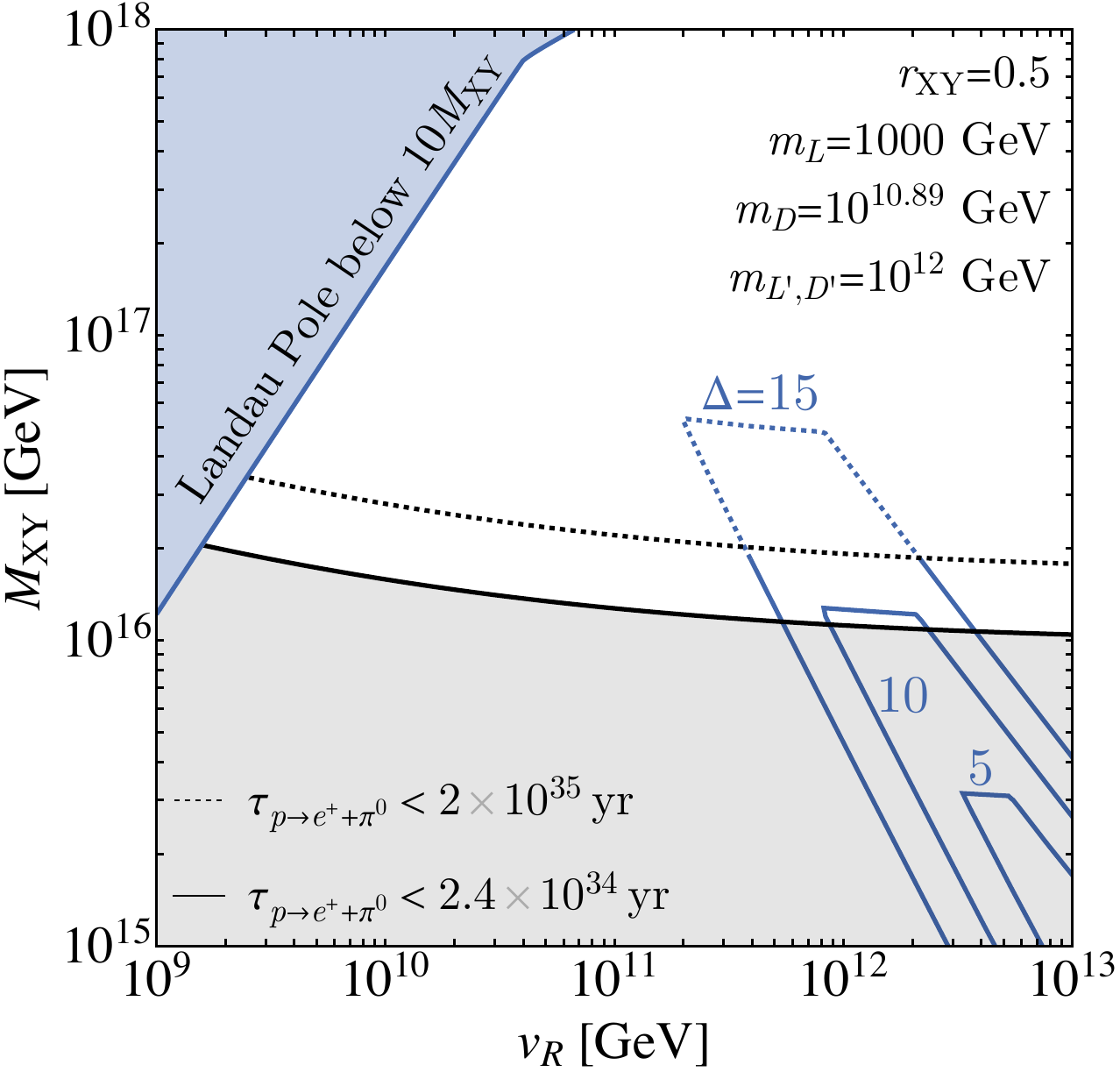}
         \caption{Lowest $m_D$ that satisfies HK constraint}
         \end{subfigure}
     \begin{subfigure}[b]{0.49\textwidth}
    \centering
    \includegraphics[width=\textwidth]{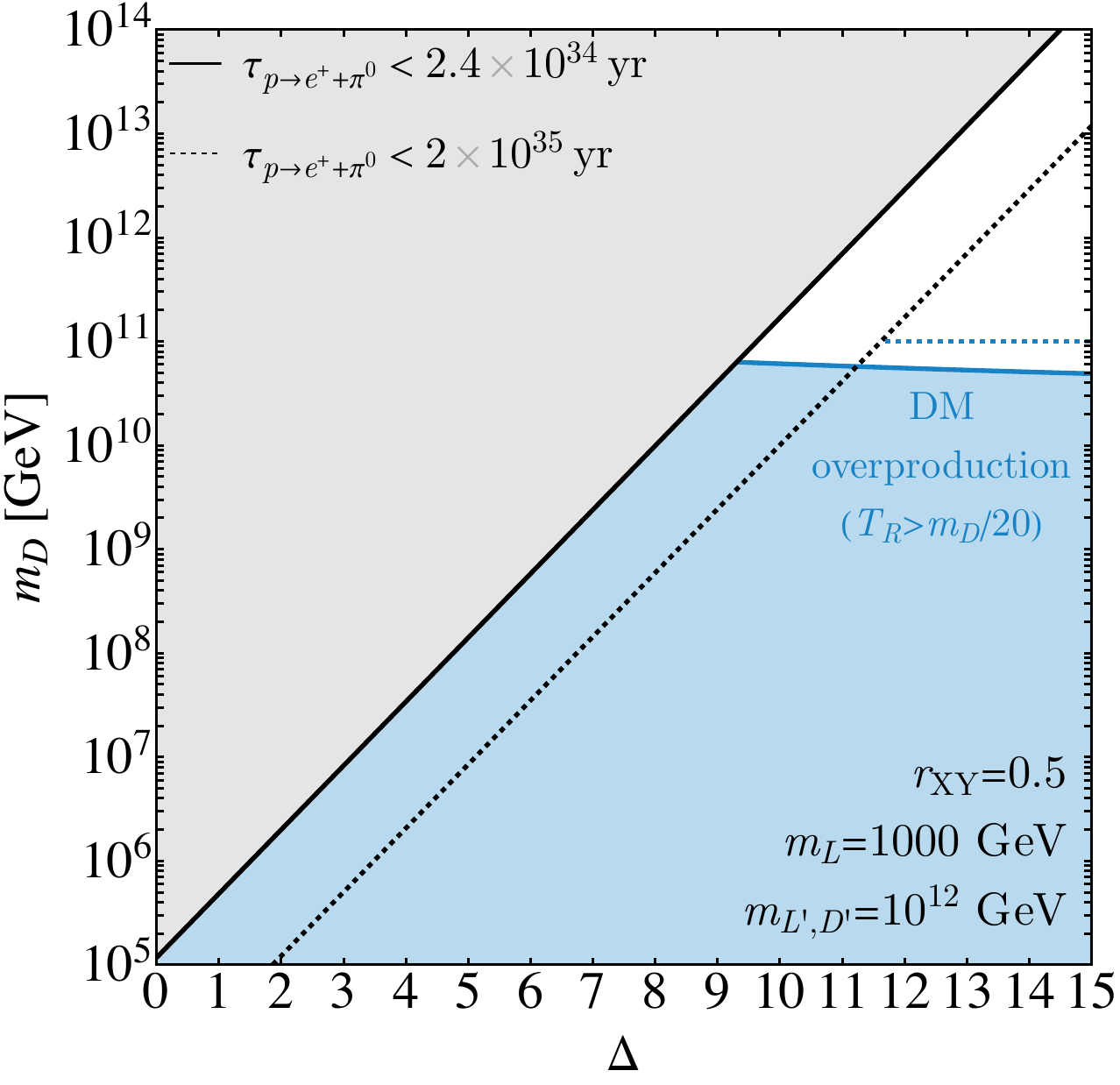}
    \caption{Constraint on $m_D$}
     \end{subfigure}
     \begin{subfigure}[b]{0.49\textwidth}
    \centering
    \includegraphics[width=\textwidth]{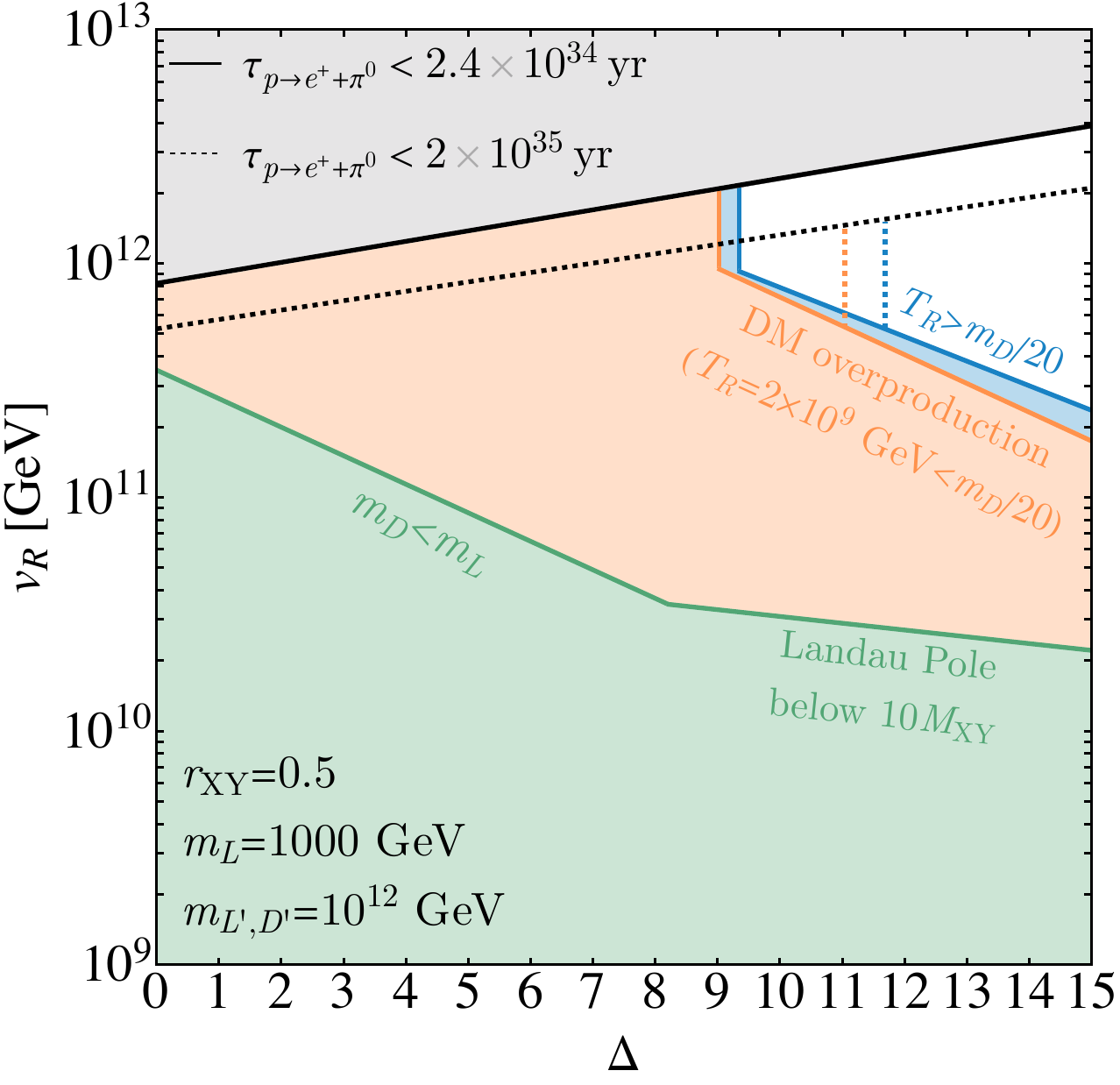}
    \caption{Constraint on $v_R$}
     \end{subfigure}

    \caption{Same as Fig.~\ref{fig:1TeVrXY2Contours} with $r_{XY}=1/2$.}
    \label{fig:1TeVrXY1o2Contours}
\end{figure*}

\begin{figure*}[h]
    \centering

 \begin{subfigure}[b]{0.49\textwidth}
         \centering
         \includegraphics[width=\textwidth]{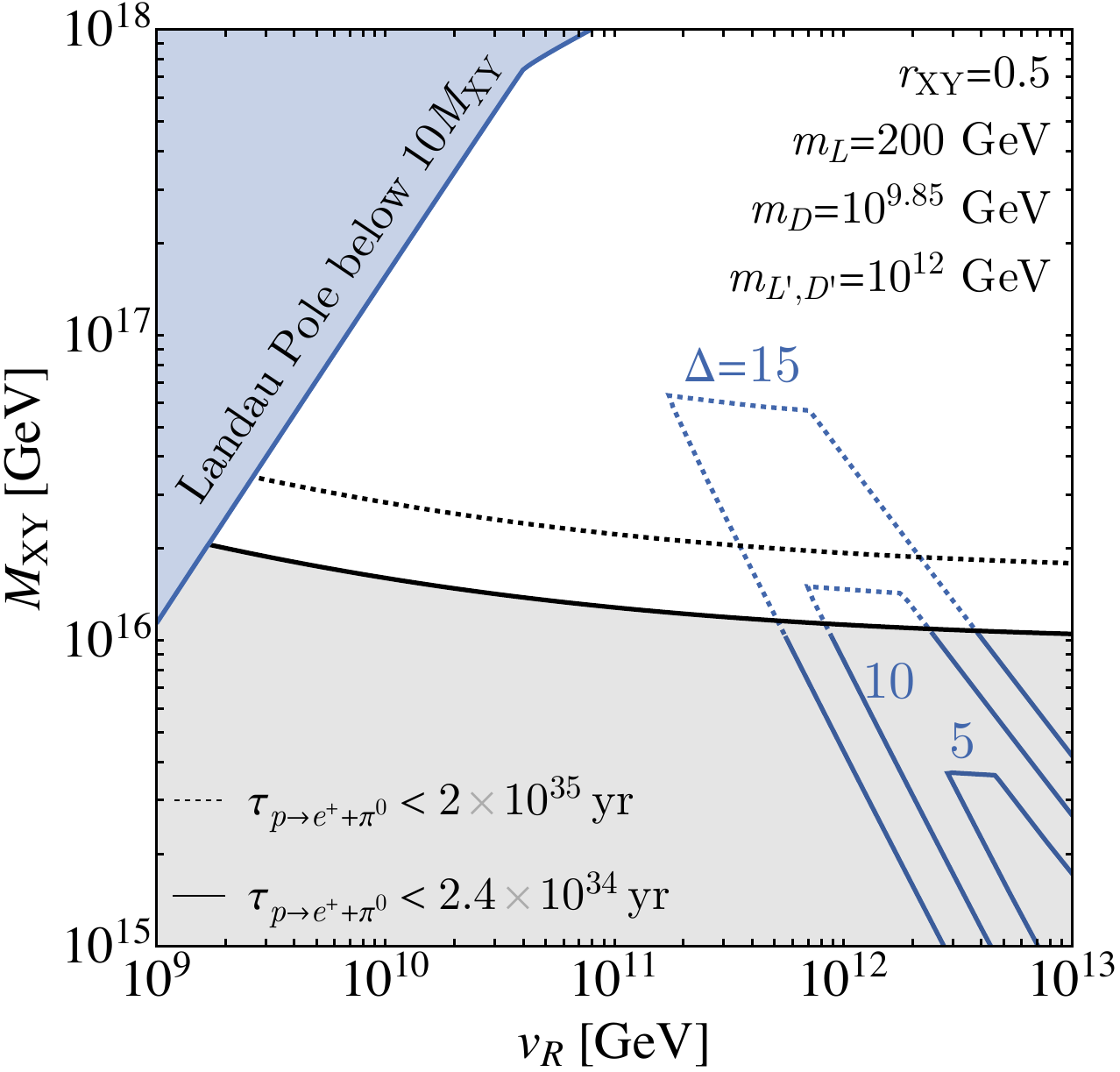}
         \caption{Lowest $m_D$ that satisfies SK constraint}
     \end{subfigure}
     \begin{subfigure}[b]{0.49\textwidth}
         \centering
         \includegraphics[width=\textwidth]{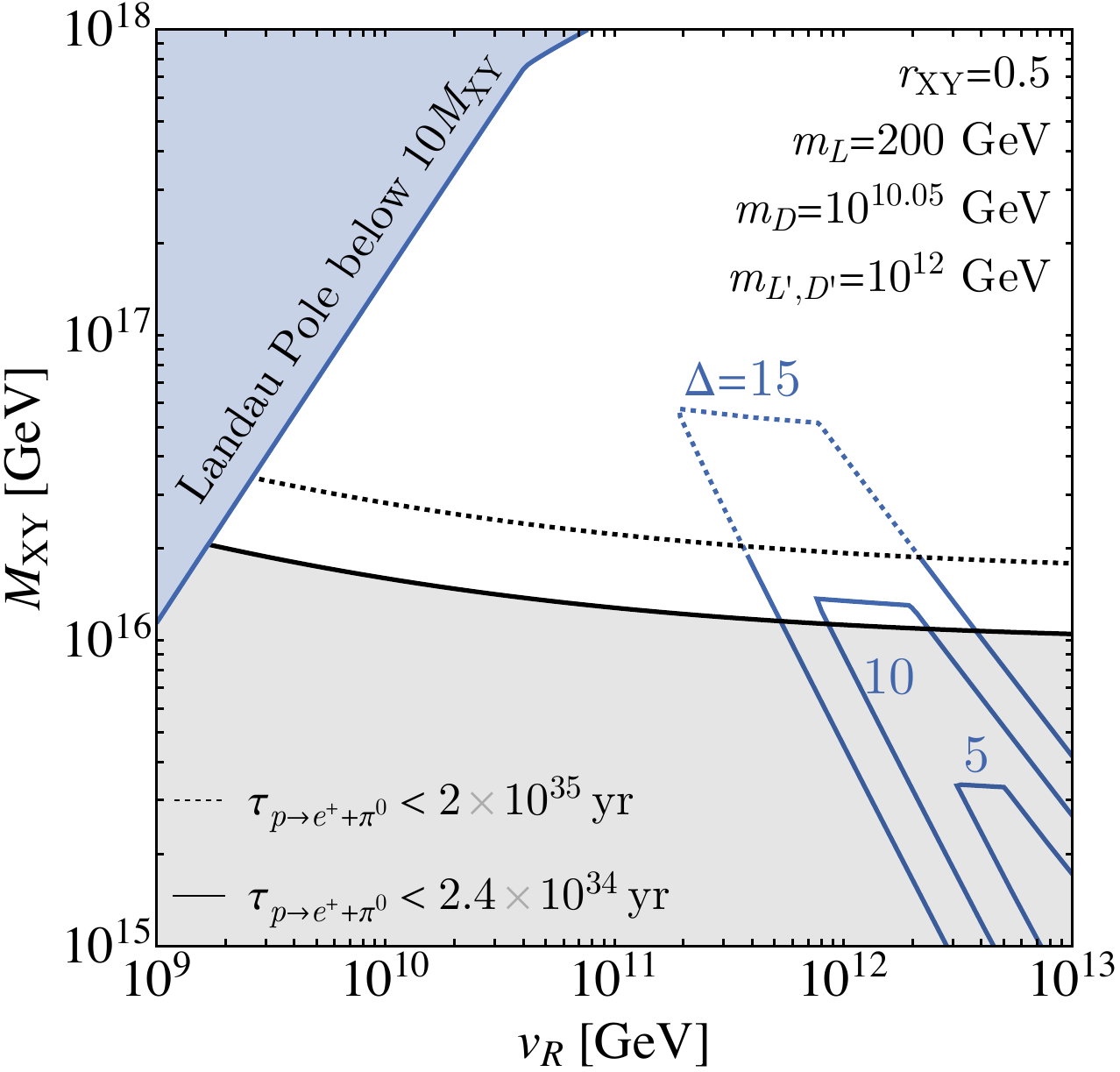}
         \caption{Lowest $m_D$ that satisfies HK constraint}
         \end{subfigure}
        \begin{subfigure}[b]{0.49\textwidth}
    \centering
    \includegraphics[width=\textwidth]{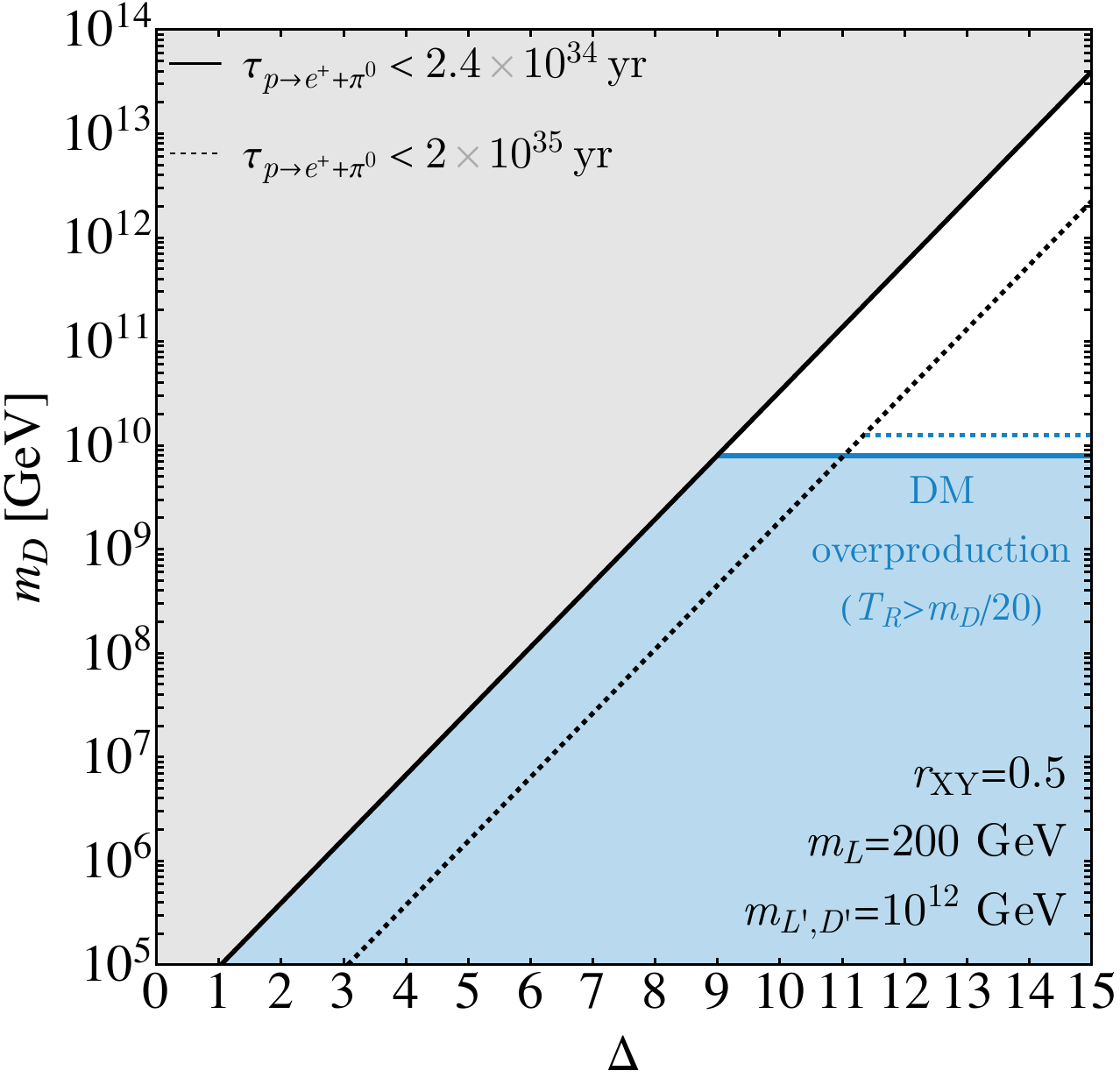}
    \caption{Constraint on $m_D$}
     \end{subfigure}
     \begin{subfigure}[b]{0.49\textwidth}
    \centering
\includegraphics[width=\textwidth]{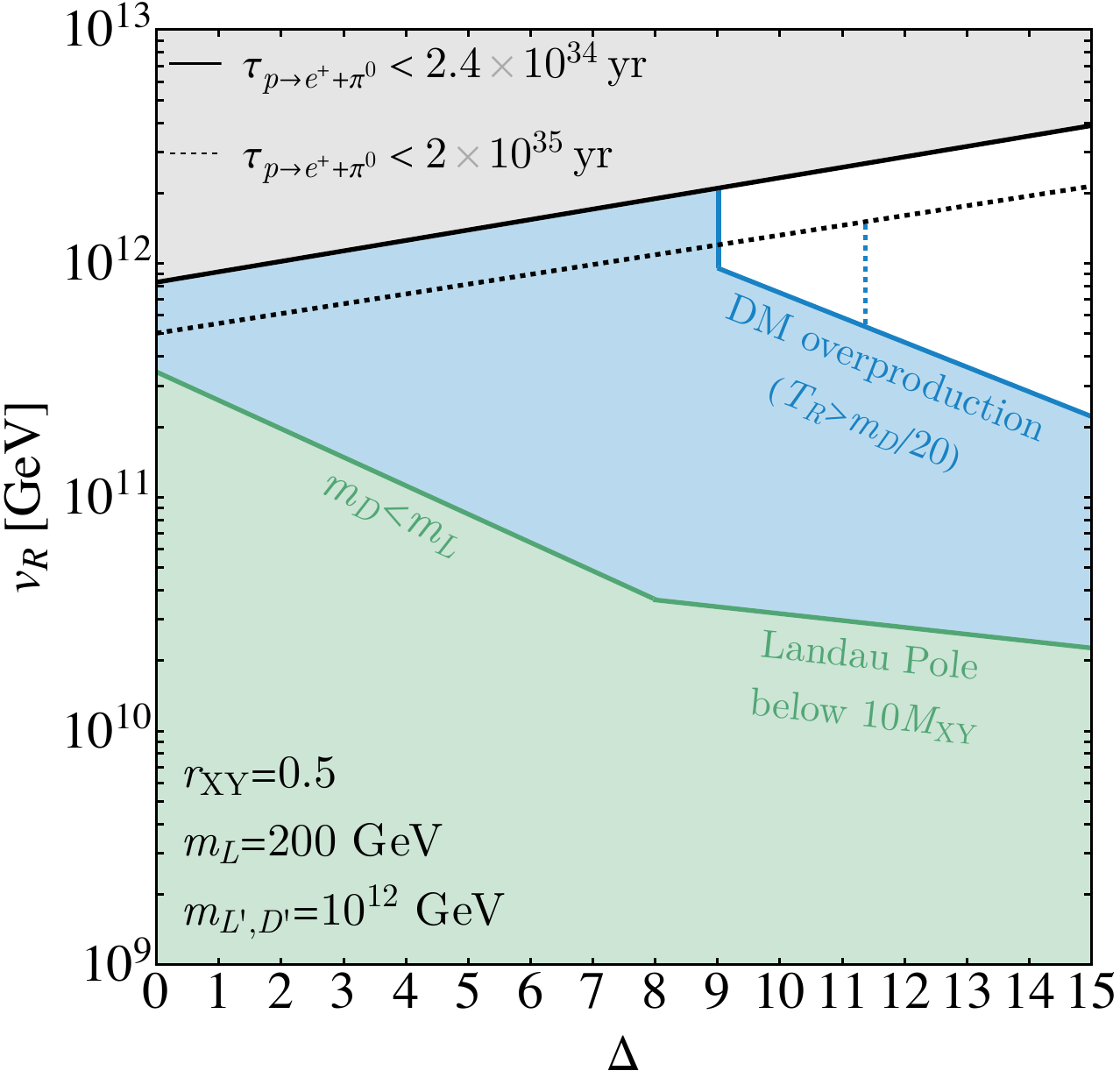}
    \caption{Constraint on $v_R$}
     \end{subfigure}

    \caption{Same as Fig.~\ref{fig:1TeVrXY2Contours} with $m_L=200$ GeV and $r_{XY}=1/2$.}
    \label{fig:100GeVrXY1o2Contours}
\end{figure*}

In this appendix, we show the constraints on $M_{XY}$, $m_D$ and $v_R$ for $r_{XY}=1/2$. Figs.~\ref{fig:1TeVrXY1o2Contours} and~\ref{fig:100GeVrXY1o2Contours} show the constraints for $m_L=1$ TeV and $200$ GeV, respectively. Because the preferred unification scale decreases, the proton decay constraint becomes stronger and the minimal $\Delta$ is larger than that for $r_{XY}=2$.

\clearpage
\bibliographystyle{JHEP}
\bibliography{biblio.bib}

\end{document}